\UseRawInputEncoding
\documentclass[onecolumn]{revtex4-2}

\usepackage{amsmath,amssymb}
\usepackage[colorlinks,
   linkcolor=blue,
  anchorcolor=blue,
  citecolor=blue]{hyperref}

\usepackage{subfigure}
\usepackage{graphicx}
\usepackage{dcolumn}
\usepackage{bm}
\usepackage[section]{placeins}
\usepackage{xcolor,float}
\usepackage{epstopdf}
\usepackage{booktabs}
\usepackage{threeparttable}
 \usepackage{multirow}
 \usepackage{booktabs}

\begin{document}

\title{Probing Bardeen-Kiselev black hole with cosmological constant caused by Einstein equations coupled with nonlinear electrodynamics using quasinormal modes and greybody bounds }

\author{S. R. Wu\footnote{gs.srwu20@gzu.edu.cn}$^{1}$, B. Q. Wang\footnote{wangbingqian13@yeah.net}$^{2}$, and Z. W. Long\footnote{zwlong@gzu.edu.cn (Corresponding author)}$^{1}$}
\affiliation{$^1$ College of Physics, Guizhou University, Guiyang, 550025, China.\\
$^{2}$ College Pharmacy, Guizhou University of Traditional Chinese Medicine, Guiyang, 550025, China.\\}

\date{\today}
 \begin{abstract}
In this work, we investigate a static and spherically symmetric Bardeen-Kiselev black hole with cosmological constant which is a solution of the Einstein-non-linear Maxwell field equations along with a quintessential field. We compute the quasinormal frequencies for Bardeen-Kiselev black hole(BH) with cosmological constant due to electromagnetic and gravitational perturbations. By varying the BH parameters, we discuss the behaviour of both real and imaginary parts of the BH quasinormal frequencies and compare frequencies with Reissner-Nordstr\"om-de Sitter BH surrounded by quintessence (RN-dSQ). Interestingly, it shows that the response of Bardeen-Kiselev BH with cosmological constant and RN-dSQ under electromagnetic perturbations are different when the charge parameter $q$, the state parameter $w$ and the normalization factor $c$ are varied, but for the gravitational perturbations, the response of Bardeen-Kiselev BH with cosmological constant and RN-dSQ are different only when the charge parameter $q$ is varied. Therefore, compared with the gravitational perturbations, the electromagnetic perturbations can be used to understand nonlinear and linear electromagnetic fields in curved spacetime separately. Another interesting observation is that due to the presence of quintessence, the electromagnetic perturbations around the Bardeen-Kiselev BH with
cosmological constant damps faster and oscillates slowly, and for the gravitational perturbations, the quasinormal mode decays slowly and oscillates slowly. We also study the reflection and transmission coefficients along with absorption cross section for the Bardeen-Kiselev BH with cosmological constant, it shows that the transmission coefficients
will increase due to the presence of quintessence.

\end{abstract}

 \maketitle
{\centering\section { Introduction}}

In the last two decades, the gravitational wave searches by the LIGO Scientific and Virgo collaboration for the merger of compact binaries \cite{ret1,ret2,ret3,ret4} and the first measurement of the shadow image captured by Event Horizon Telescope \cite{ret5,ret6,ret7,ret8,ret9,ret10} have provided convincing evidences in terms of the existence of BHs in the center
of many galaxies, such as the elliptical M87 galaxy and spiral Milky Way\cite {ret11}. Actually, the concept of BH exactly originates from one of the most important predictions of general relativity, and in view of their causal structure, BHs have a surface to which any wave or particle passing through this surface can not return, this surface is labled as the event horizon. General relativity is a theory which is plagued with the existence of singularities, in particular, gravitational singularities appear in general relativity in the context of black holes, i.e. some places in spacetime where the geodesics are interrupted \cite{ret12}. Therefore, how to avoide the singularities in general relativity is one of our most pressing problems. Possibly, the regular BHs are supposed one of the alternatives to solve the problem of the existence of singularities, the so called ``regular BHs" means that the BHs do not have a spacetime singularity at the origin. Usually, one can obtain a regular BH solution is coupling general relativity to those described by the nonlinear electrodynamics (NED), one of the interesting properties of the NED is that it could eliminate the curvature singularity from the BH solutions \cite{retzy,retka}, and NED can be considered to generalize Maxwell's theory to strong field regimes, naturallly, this theory can provide a choice for studying charged black holes where we deal with strong electromagnetic and gravitational fields\cite{retRAM}.

Regarding the theory of NED coupling to the Einstein equations, it should be noted that we need to find a suitable gauge invariant Lagrangian and its energy-momentum tensor. Bardeen derived the first solution of regular black holes with non-singular geometry which satisfies the weak energy condition \cite{ret13}. Noted that this solution is not a vacuum solution rather gravity is corrected by introducing some matter, that is known as energy momentum tensor which is introduced in the Einstein¡¯s equation. Afterwards, researches show that the energy momentum tensor necessary to derive the regular BH solution is essentially the gravitational field of some magnetic monopole, which originates from a specific form of non-linear electrodynamics \cite{ret14}, and the latest work showed that the Kiselev solution becomes an exact solution of the Einstein-power-Maxwell equations (with or without cosmological constant) using either an ansatz involving electric charges and fields, or a magnetic monopole ansatz\cite{retMAD}.
 Another alternative that avoids the presence of singularities is the black bounce \cite{retA}, this type of solution has a minimal nonzero area, and in the context of general relativity, this solution is not described only by coupling with nonlinear electrodynamics, but also requires an additional form of matter \cite{retE}. Of course, as the research progresses, other forms of solutions have been proposed thereafter \cite{ret15,ret16,ret17}.

The stability of a BH spacetime is probably one of the most interssting issues in the context of general relativity, since a BH under certain perturbations can help to study the nature of the BH itself. In general, by studying the BH merger or the field evolution of the BH background, one can discuss the stability of BH spacetime under perturbation \cite{retSA,retSR}. It is well know that a BH in perturbation can emit gravitational waves which are dominated by quasinormal mode (QNM), and this mode is a complex frequency of excited oscillation mode \cite{ret18}, where the real part and the imaginary part of QNMs frequency are the oscillation frequency and damping of the BH respectively.
For electrically or magnetically charged BHs, the Einstein-Maxwell theory is usually considered to be the standard theory and leads to the well known Reissner-Nordstr\"om solution\cite{retEWL}. Noted that the QNMs may be subject to corrections due to some effects of electrodynamics beyond the Maxwell theory, and which actually exist at least when we consider the quantum electrodynamics. Based on these considerations, we are interested to discuss the QNMs of charged BHs in a framework beyond the standard Einstein-Maxwell theory.
Usually, a BH in perturbation can be classified into three distinct stages, the first one consists of an initial outburst of wave which depends completely on the initial perturbing field, the second one is a usually long period of damping proper oscillations which dominated by the QNM, this stage contributes to gravitational wave detection, the final stage is a power law tail behaviour at very late times. In this work, we focus on the second stage of the evolution of perturbations represented by QNM.
Generally the beginning of QNMs calculation is to reduce the perturbation equations into a 2 dimensional wavelike form with decoupled angular variables, and if the variables are decoupled, the equation for time and radial variables will transform into a Schr\"odinger-like form in a stationary background. Then the corresponding potential function can be determined, which is exactly the key for the computation of the QNM frequencies. The QNM can be calculated by various methods, such as continued fraction method \cite{ret19}, WKB method \cite{ret20}, P\"oschl-Teller approximation \cite{ret21}, the time domain method \cite{retEN} and the expansion method \cite{retSR}. In view of the accuracy of the WKB method in terms of the real and the imaginary parts of the dominant QNMs with $n \leq l$, we use this method to the QNM frequencies calculation. In our work we focus on the behavior of the dynamical response of the spherically symmetric and magnetical BH, which represents the exact solutions of coupled Einstein gravity and NED to small electromagnetic perturbations and gravitational perturbations, especially, we intend to identify whether it would distinguish the BHs related to the NED from those BHs related to the standard linear electrodynamics due to their response to electromagnetic perturbations and gravitational perturbations. Perturbations of BHs imply the exploration of stability of their spacetime, and the stability of different BHs involved NED has been considered in \cite{retNB, retNBS}. In addition, outside the event horizon, the effective potential plays a role of filter \cite{ret22}. Concretely speaking, some waves probably pass through the effective potential and transmit to infinity, while some waves are probably reflected back by the effective potential, thus if we consider an observer at infinity, obviously the radiation received by the observer is different from the radiation emitted from the event horizon \cite{ret23}. This behavior in Hawking radiation is known as the greybody factors \cite{ret24}, which can encode information about the horizon structure of BHs theoretically and modify the quasinormal spectra experimentally \cite{retPK}, also in order to estimate the transmission probability of radiations from a BH¡¯s event horizon to its asymptotic region, it is necessary to study the graybody factors of perturbations.
It is well known that BHs in real world are not isolated and are not embedded in empty backgrounds. In astronomy, stellar BHs are expected to be surrounded by matter or fluids, especially the solutions of black holes gotten by Kiselev just describe such spacetimes surrounded by anisotropic fluids\cite{retVVK}. Moreover, the Kiselev solution can be implemented for more generic backgrounds of radiation, dust, quintessence, cosmological constant and phantom fields as well as for any realistic combination of these cosmological fields. With this in mind, one may intend to discuss some fascinating facts such as what are the influences of these surrounding fields on the features and behaviors of BHs or how BHs affect these cosmological surrounding fields and what are the consequences\cite{retYHE}.
Originally the fluid surrounding the BH was called quintessence, whose equation of state is given by $p=w\rho_{q}$, with $p$ is the pressure, $\rho_{q}$ is the density of energy and $w$ is the state parameter. While according to the Ref.\cite{retMVI}, the appropriate the correct interpretation corresponds to a mix of fluids. It should be noted that the resulting anisotropic fluid surrounding this kind of BHs could be mimicked by a composition of an electrically charged fluid and an electromagnetic and/or scalar field, whose form depends on the radial coordinate $r$ in this multicomponent model\cite{retBCU}. The study of quasinormal modes of BHs surrounded by quintessence draw much attention. The effects of the quintessential parameter on quasinormal frequencies of different BH spacetimes were widely discussed in \cite{retSCH,retJPM,retMSA}.

The authors in Ref.\cite{retMAN} discussed the Bardeen solution with a cosmological constant surrounded by quintessence, they show that this solution can be obtained by Einstein equations coupled with nonlinear electrodynamics and it is not always regular and what the conditions for regularity are, also they analyzed the thermodynamics associated with this type of solution by establishing the form of the Smarr formula and the first law of thermodynamics.  In this work, we study the stability of this solution and the QNMs of perturbations, also the reflection coefficient, gray-body factor and absorption coefficient of perturbations are invovled. This paper is organized as follows. In Sec. \ref{secII}, we present a brief review on Bardeen-Kiselev BH with cosmological constant. In Sec. \ref{secIII}, the gravitational perturbations and electromagnetic perturbations for Bardeen-Kiselev BH with cosmological constant are reported. The QNMs of the Bardeen-Kiselev BH with cosmological constant for gravitational perturbations and electromagnetic perturbations are analyzed and compare frequencies with RN-dSQ \cite{HLI} in Sec. \ref{secIV}. In Sec. \ref{secV}, we focus on a discussion about the greybody factor, the reflection coefficient and the total absorption cross section. Finally, the work is summarized in Sec. \ref{secVI}. In this paper we consider natural units, where $c=G=\hbar=1$, and the metric signature $(+,-,-,-)$.

{\centering  \section{A brief review on Bardeen-Kiselev black hole with cosmological constant} \label{secII} }

This section deals with a very brief introduction to the Bardeen-Kiselev BH following the work in Ref.\cite{retMAN}, the corresponding action is given by

\begin{equation}  S=\int d^{4} x \sqrt{-g}[R+2 \lambda+L(F)],             \label{eq1} \end{equation}
here $R$, $\lambda$ and $L(F)=\frac{24 \sqrt{2} M q^{2}}{8\pi\left(\sqrt{\frac{2 q^{2}}{F}}+2 q^{2}\right)^{\frac{5}{2}}}-\frac{6 w c\left(\frac{2 F}{q^{2}}\right)^{\frac{3}{4}(w+1)}}{8\pi}$ denote the scalar curvature, the cosmological constant and the function of the field strength $F$ of the nonlinear Lagrangian of electromagnetic theory respectively, here $c$ is the normalization constant related to the density of quintessence $\rho_{q}=-\frac{c}{2}\frac{3w}{r^{w+3}}$, $M$ is the total mass of the BH and $q$ is the magnetic charge. The field strength $F$ is given by
\begin{equation}  F=F^{\mu v} F_{\mu v},             \label{eq2} \end{equation}
where the scalar $F$ is $F=\frac{2 q^{2}}{r^{4}}$ and $F_{\mu \nu}$ is the electrodynamic field tensor that can be expressed in terms of a gauge potential as $F_{\mu \nu}=\partial_{\mu} A_{\nu}-\partial_{\nu} A_{\mu}$. The electromagnetic field tensor implies that $F_{\mu \nu}$ is anti-symmetric and it has only six independent components, the only nonzero component of $F_{\mu \nu}$ for a spherically symmetric spacetime that is only magnetically charged is $F_{23}=qsin\theta$.

Here one can express the covariant equations of motion as
\begin{equation}
\begin{aligned}
 & R_{\mu v}-\frac{1}{2} g_{\mu v} R+\lambda g_{\mu v}=8 \pi T_{\mu v}, \\  &\nabla_{v}\left(L_{F} F^{\mu v}\right)=0, \label{eq3}
\end{aligned}   \end{equation}
if we consider the stress-energy tensor to nonlinear electrodynamics
\begin{equation} T_{\mu v}=g_{\mu v} L(F)-L_{F} F_{\mu}^{\alpha} F_{v \alpha},             \label{eq4} \end{equation}
with $L_{F}=\frac{\partial L(F)}{\partial F}$, then we will look for solutions which are static and spherically symmetric for Einstein equation
\begin{equation}  d s^{2}=f(r) d t^{2}-\frac{1}{f(r)} d r^{2}-r^{2}\left(d \theta^{2}+\sin ^{2} \theta d \phi^{2}\right),             \label{eq5} \end{equation}
where
\begin{equation}  f(r)=1-\frac{2 c}{r^{3 w+1}}-\frac{2 M r^{2}}{\left(q^{2}+r^{2}\right)^{\frac{3}{2}}}-\frac{\lambda r^{2}}{3},    \label{eq6} \end{equation}
when $c>0$ and $-1\leq w \leq0$, the anisotropic fluid fulfills the null energy condition\cite{retPBO}. Nonetheless, in this work we will focus on the cases $-1\leq w \leq-1/3$ since we are interested in asymptotically dS-like spacetimes. The cases $-1/3\leq w \leq0$ correspond to asymptotically flat BHs, which remain out of the scope of this work. And regarding the Eq.\eqref{eq6}, we find the Kiselev-(anti-)de Sitter solution with cosmological constant for the limit $q$ $\rightarrow$ $0$ and the Bardeen-(anti-)de Sitter solution for $c$ $\rightarrow$ $0$\cite{retBM}.

{\centering  \section{Gravitational perturbations and electromagnetic perturbations} \label{secIII} }

The study of BH perturbations was first proposed by Regge and Wheeler in terms of the investigation for the odd parity type of the spherical harmonics\cite{ret29}, afterwards Zerilli generalized it to the even parity case\cite{ret30}. It is well known that generally there are two different kinds of perturbations of BHs in the context of the general theory of relativity: the test field in a BH background and the perturbation of the metric. The former one is achieved through the method of solving the dynamical equation for the given test field in the background of the BH, and the second one is obtained by linearising the Einstein equation to derived the evolution equations, i.e. the gravitational perturbation. In view of the fact that comparison with the strength of the external fields decaying in the vicinity of BH, the gravitational radiation is the strongest one. In this section, we firstly focus on the gravitational perturbation of the Bardeen-Kiselev BH with cosmological constant. The general procedure for gravitational perturbation is to introduce a small perturbation $(h_{\mu v}\ll1)$ into the static background metric $(\tilde{g}_{\mu v})$, then we assume the perturbed background metric $g_{\mu v}$ as

\begin{equation}  g_{\mu v}=\tilde{g}_{\mu v}+h_{\mu v},             \label{eq7} \end{equation}

The perturbations $h_{\mu v}$ can be decomposed as

\begin{equation}
h_{\mu v}=\left(\begin{array}{cccc}0 & 0 & 0 & h_{0}(t, r) \\ 0 & 0 & 0 & h_{1}(t, r) \\ 0 & 0 & 0 & 0 \\ h_{0}(t, r) & h_{1}(t, r) & 0 & 0\end{array}\right)(\xi(\theta)), \label{eq8}
\end{equation}
this formalism is similar to the axial decomposition in the Regge-Wheeler gauge. Noted that it is not exactly the same gauge, we prefer to obtain the form of $\xi(\theta)$ through the
field equations rather than by imposition of some kind of expansion in terms of spherical harmonics \cite{retSCU}.

Here we have taken into account the perturbation in the energy momentum tensor and obtain the Einstein's equation as

\begin{equation}  G_{\mu v}+\lambda g_{\mu v}=8\pi T_{\mu v},             \label{eq9} \end{equation}
and
\begin{equation}
\begin{aligned}
&\delta R_{\mu v}=-\nabla_{\alpha} \delta \Gamma_{\mu v}^{\alpha}+\nabla_{v} \delta \Gamma_{\mu \alpha}^{\alpha} \\
&\delta \Gamma_{\beta \gamma}^{\alpha}=\frac{1}{2} \tilde{g}^{\alpha v}\left(\partial_{\gamma} h_{\beta v}+\partial_{\beta} h_{\gamma v}-\partial_{v} h_{\beta \gamma}\right),
 \label{eq10}      \end{aligned}      \end{equation}
it should be noted that the Ref.\cite{retJPM} studied the quasinormal modes of gravitational perturbation around regular Bardeen BH surrounded by quintessence in vacuum regardless of  perturbed energy momentum tensor, and we may view that our consideration is more interesting and general.
The perturbed Einstein equations then lead to the equalities as Eq.\eqref{eq11} with $T_{\theta\phi}=0$ and $T_{r\phi}=h_{1} L \xi(\theta)$
\begin{equation}
\begin{aligned}
&-\frac{1}{f} \frac{\partial}{\partial t} h_{0}+\frac{\partial}{\partial r}\left(f h_{1}\right)=0, \\
&\frac{1}{f}\left(\frac{\partial^{2}}{\partial t^{2}} h_{1}-\frac{\partial{ }^{2} h_{0}}{\partial t \partial r}+\frac{2}{r} \frac{\partial}{\partial t} h_{0}\right)+\left[\frac{\ell(\ell+1)-2}{r^{2}}+\frac{2}{r} f^{\prime}+f^{\prime \prime}+2\left(k^{2} L+\lambda\right)\right] h_{1}=0\\
&\frac{\partial^{2}\xi}{\partial \theta^{2}}-\frac{cos\theta}{sin\theta}\frac{\partial\xi}{\partial \theta}+l(l+1)\xi=0,       \label{eq11}      \end{aligned}      \end{equation}
here $l$ is the multipole number, $k^{2}=8\pi$ and $\xi(\theta)=P_{\ell}(\cos \theta)$(the Legendre Polynomials).
By using the definition $\varphi(t, r)=\frac{f}{r} h_{1}(t, r)$, Eq.\eqref{eq11} can be expressed as
\begin{equation}  {\left[\frac{\partial^{2}}{\partial t^{2}}-f^{2} \frac{\partial^{2}}{\partial r^{2}}-f f^{\prime} \frac{\partial}{\partial r}+f\left(\frac{\ell(\ell+1)+2(f-1)}{r^{2}}+\frac{f^{\prime}}{r}+f^{\prime \prime}+2\left(k^{2} L+\lambda\right)\right)\right] \varphi(t, r)=0},  \label{eq12} \end{equation}
where $\prime$ denotes derivative with respect to the radial coordinate $r$.
In order to facilitate this procedure, we change the variable $r$ to the tortoise coordinate $r_{*}$ with the defination $d r=f(r) d r_{*}$, and when $r \to r_{c}$, $r_{*}\to \infty$, when $r \to r_{+}$, $r_{*}\to -\infty$, with $r_{c}$ is the cosmological horizon and $r_{+}$ is the event horizon, then we have
 \begin{equation}  {\left[\frac{\partial^{2}}{\partial t^{2}}-\frac{\partial^{2}}{\partial r_{*}^{2}}+V(r)\right] \varphi\left(t, r_{*}\right)=0 },  \label{eq13} \end{equation}
with the effective potential
 \begin{equation}  V(r)=f\left(\frac{\ell(\ell+1)+2(f-1)}{r^{2}}+\frac{f^{\prime}}{r}+f^{\prime \prime}+2\left(k^{2} L+\lambda\right)\right),  \label{eq14} \end{equation}
 this formalism is same as was found in Ref.\cite{retBTO} in the limit $\lambda\to0$, but there is a difference in numerical factor in the effective potential in terms of the coefficient of $L$ and $f$.

Here if we consider the temporal dependence as $\varphi \sim \Psi e^{-i{\omega}t}$, we can get the master equation for gravitational perturbations of the BH as
\begin{equation}\frac{d^{2}\Psi (r_{*}) }{dr^{2}_{*}}+(\omega^{2}-V(r))\Psi(r_{*})=0, \label{eq15} \end{equation}
 the parameter $\omega$ denotes the dissipative modes in time, thus the BH will oscillate after the perturbation and then go back to a stable state, which is known as quasinormal frequencies.

Next, the behaviour of the dynamical response of the spherically symmetric BH for electromagnetic perturbations is our topic. We decompose 4-vector potential of the electromagnetic field as unperturbed background potential $\widetilde{A}_{\mu}$ and perturbed part $\delta A_{\mu}$
\begin{equation}A_{\mu}=\widetilde{A}_{\mu}+\delta A_{\mu},\label{eq16} \end{equation}
in view of the static and spherically symmetric background, we consider the unperturbed 4-vector potential of magnetically charged BH as
\begin{equation}\widetilde{A}_{\mu}=-q \cos \theta \delta_{\mu}^{\phi}.\label{eq17} \end{equation}
In spherically symmetric background, the perturbation in vector potential can be expressed as a superposition of vector spherical harmonics, i.e.
\begin{equation}\delta A_{\mu}=\sum_{\ell, m}\left[\begin{array}{c}0 \\ 0 \\ \psi(t, r) \frac{\partial_{\phi} Y_{\ell m}(\theta, \phi)}{\sin \theta} \\ -\psi(t, r) \sin \theta \partial_{\theta} Y_{\ell m}(\theta, \phi)\end{array}\right],\label{eq18} \end{equation}
with $Y_{\ell m}(\theta, \phi)$ denotes scalar spherical harmonics.

The nonvanishing covariant components of the electromagnetic field tensor of the 4-potential Eq.\eqref{eq16} with perturbation Eq.\eqref{eq18} are given by
\begin{equation}
\begin{aligned}
&F_{t \theta}=\frac{1}{\sin \theta} \partial_{t} \psi(t, r) \partial_{\phi} Y_{\ell m}(\theta, \phi), \\
&F_{t \phi}=-\sin \theta \partial_{t} \psi(t, r) \partial_{\theta} Y_{\ell m}(\theta, \phi), \\
&F_{r \theta}=\frac{1}{\sin \theta} \partial_{r} \psi(t, r) \partial_{\phi} Y_{\ell m}(\theta, \phi), \\
&F_{r \phi}=-\sin \theta \partial_{r} \psi(t, r) \partial_{\theta} Y_{\ell m}(\theta, \phi),    \\
&F_{\theta \phi}=\sin \theta\left(q+\ell(\ell+1) \psi(t, r) Y_{\ell m}(\theta, \phi)\right),
 \label{eq19}      \end{aligned}      \end{equation}
by the relation $F^{\mu \nu}=g^{\mu \alpha} g^{\nu \beta} F_{\alpha \beta}$, all non zero contravariant components of electromagnetic field tensor are following:

\begin{equation}
\begin{aligned}
&F^{t \theta}=-\frac{1}{f(r) r^{2} \sin \theta} \partial_{t} \psi(t, r) \partial_{\phi} Y_{\ell m}(\theta, \phi), \\
&F^{t \phi}=\frac{1}{f(r) r^{2} \sin \theta} \partial_{t} \psi(t, r) \partial_{\theta} Y_{\ell m}(\theta, \phi),  \\
&F^{r \theta}=\frac{f(r)}{r^{2} \sin \theta} \partial_{r} \psi(t, r) \partial_{\phi} Y_{\ell m}(\theta, \phi),  \\
&F^{r \phi}=-\frac{f(r)}{r^{2} \sin \theta} \partial_{r} \psi(t, r) \partial_{\theta} Y_{\ell m}(\theta, \phi),   \\
&F^{\theta \phi}=\frac{1}{r^{4} \sin \theta}\left(q+\ell(\ell+1) \psi(t, r) Y_{\ell m}(\theta, \phi)\right).
 \label{eq20}      \end{aligned}      \end{equation}
In view of the fact that the field strength remains same at zeroth order but has components in first order for the total 4-vector potential, combining Eq.\eqref{eq19} and Eq.\eqref{eq20}, we present the electromagnetic field strength $F$ as
\begin{equation} {F}\approx \frac{2 q^{2}}{r^{4}}+\frac{4 q}{r^{4}} \ell(\ell+1) \psi(t, r) Y_{\ell m}(\theta, \varphi),        \label{eq21} \end{equation}
here the first term corresponds to the unperturbed electromagnetic field strength $\tilde{F}$, and the second term represents the contribution of the perturbation to the field strength $\delta F$, which leads to $F=\tilde{F}+\delta F$.

Near $\tilde{F}$, by using Taylor series up to first order term for expanding $L_{F}$, we have
\begin{equation}
 L_{F} \approx \tilde{L}_{\tilde{F}}(\tilde{F})+\tilde{L}_{\tilde{F} \tilde{F}} \delta F,  \label{eq22}   \end{equation}
where $\tilde{L}_{\tilde{F} \tilde{F}}=\partial_{\tilde{F}}^{2}\tilde{L}=\partial_{\tilde{F}}\tilde{L}_{\tilde{F}}$, note that $\tilde{F}$ and $\tilde{L}_{\tilde{F}}$ depend on $r$.
By combining Eq.\eqref{eq20} into Eq.\eqref{eq3} we have
\begin{equation}
 \frac{\partial\left(L_{F} F^{\mu t}\right)}{\partial t}+\frac{1}{r^{2}} \frac{\partial\left(r^{2} L_{F} F^{\mu r}\right)}{\partial r}+\frac{1}{\sin \theta} \frac{\partial\left(\sin \theta L_{F} F^{\mu \theta}\right)}{\partial \theta}+\frac{\partial\left(L_{F} F^{\mu \phi}\right)}{\partial \phi}=0.  \label{eq23}   \end{equation}
For $u=\theta$ and $u=\phi$, the above equation transforms into
\begin{equation}
[-\frac{\partial^{2}}{\partial t^{2}}+(\frac{f(r)^{2}}{\tilde{L}_{\tilde{F}}} \tilde{L}_{\tilde{F}}^{\prime}+f(r) f^{\prime}(r) ) \frac{\partial}{\partial r}+f(r)^{2} \frac{\partial^{2}}{\partial r^{2}}-\frac{\ell(\ell+1)}{r^{2}} f(r) (1+\frac{4 q^{2} \tilde{L}_{\tilde{F} \tilde{F}}}{r^{4} \tilde{L}_{\tilde{F}}})] \psi(t, r)=0,
\label{eq24}   \end{equation}
where $\tilde{L}_{\tilde{F}}^{\prime}$ denotes the first derivative of $\tilde{L}_{\tilde{F}}$ with respect to $r$, for the sake of convenience, we take a transformation $\psi(t, r)=\left(\widetilde{L}_{\tilde{F}}\right)^{-\frac{1}{2}} \varphi(t, r)$, and considering the tortoise coordinate, we obtain the Schr\"odinger-like wave equation
\begin{equation}
\left[\frac{\partial^{2}}{\partial t^{2}}-\frac{\partial^{2}}{\partial r_{*}^{2}}+V(r)\right] \varphi\left(t, r_{*}\right)=0,    \label{eq25}     \end{equation}
with the effective potential
\begin{equation}
 V(r)=-\frac{f^{2}}{4}\left(\frac{\widetilde{L}_{\tilde{F}}^{\prime}}{\widetilde{L}_{\tilde{F}}}\right)^{2}+\frac{f f^{\prime}}{2} \frac{\tilde{L}_{\tilde{F}}^{\prime}}{\widetilde{L}_{\tilde{F}}}+\frac{f^{2}}{2} \frac{\widetilde{L}_{\tilde{F}}^{\prime \prime}}{\widetilde{L}_{\tilde{F}}}+\frac{\ell(\ell+1)}{r^{2}} f(r)\left(1+\frac{4 q^{2} \widetilde{L}_{\tilde{F} \tilde{F}}}{r^{4} \widetilde{L}_{\tilde{F}}}\right),              \label{eq26}         \end{equation}
this formalism is same as were found in Ref.\cite{retSA} and Ref.\cite{retBTOS}, also there is a difference in numerical factor in the effective potential.

\begin{figure}[htbp]
\begin{tabular}{cc}
\begin{minipage}[t]{0.45\linewidth}
\centerline{\includegraphics[width=7.0cm]{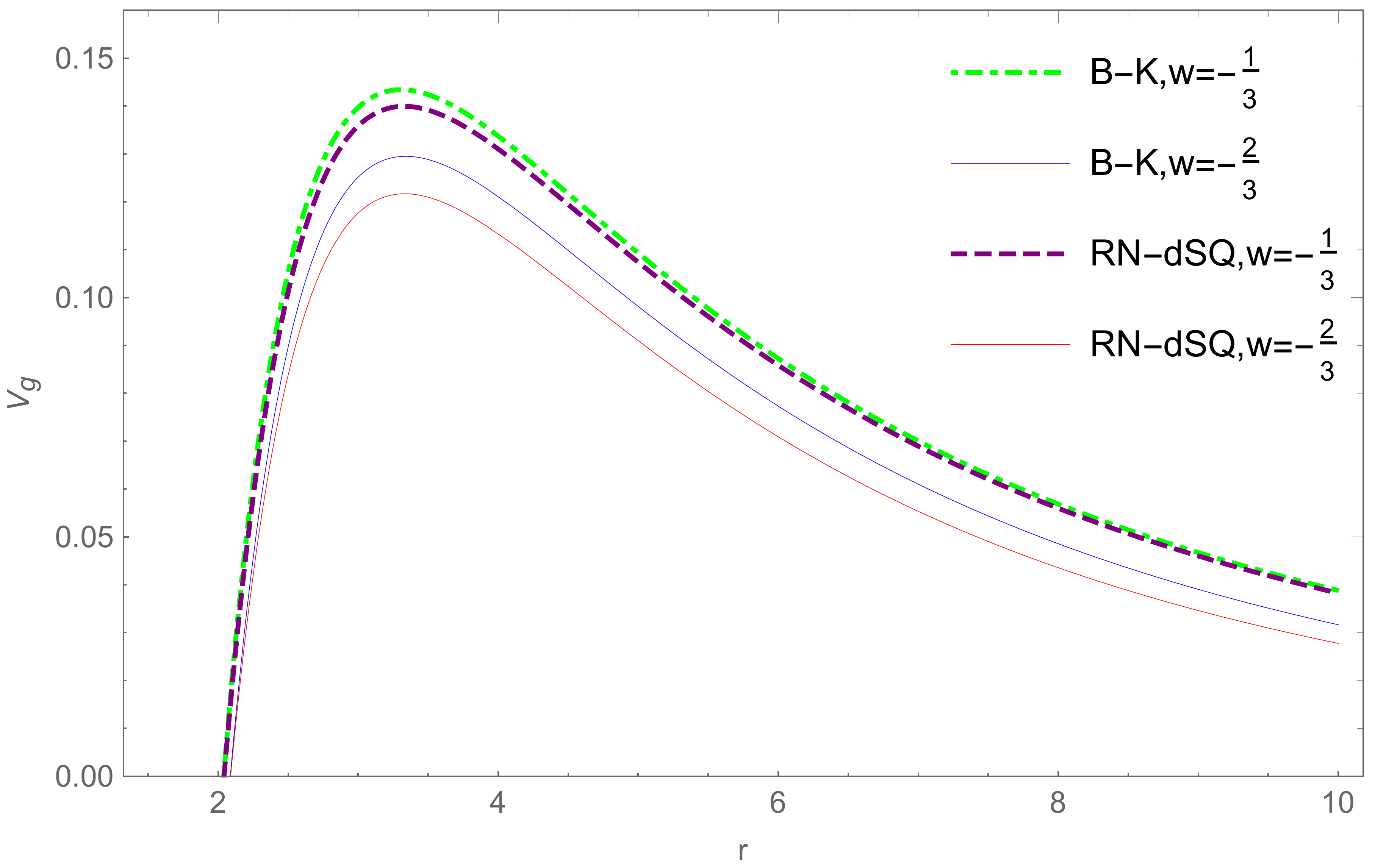}}
\centerline{(a)}
\end{minipage}
\hspace{7mm}
\begin{minipage}[t]{0.45\linewidth}
\centerline{\includegraphics[width=7.0cm]{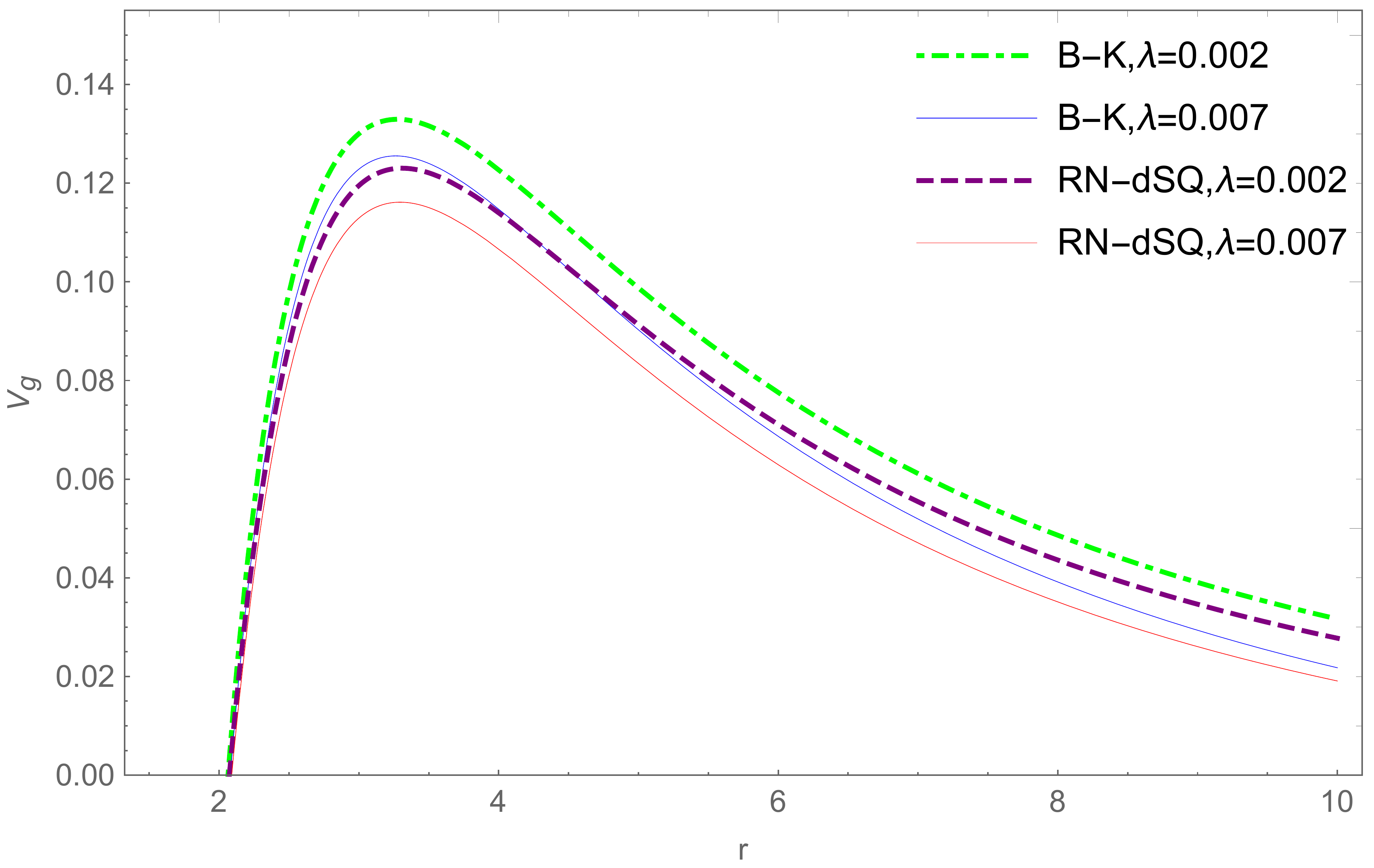}}
\centerline{(b)}
\end{minipage}
\\
\begin{minipage}[t]{0.45\linewidth}
\centerline{\includegraphics[width=7.0cm]{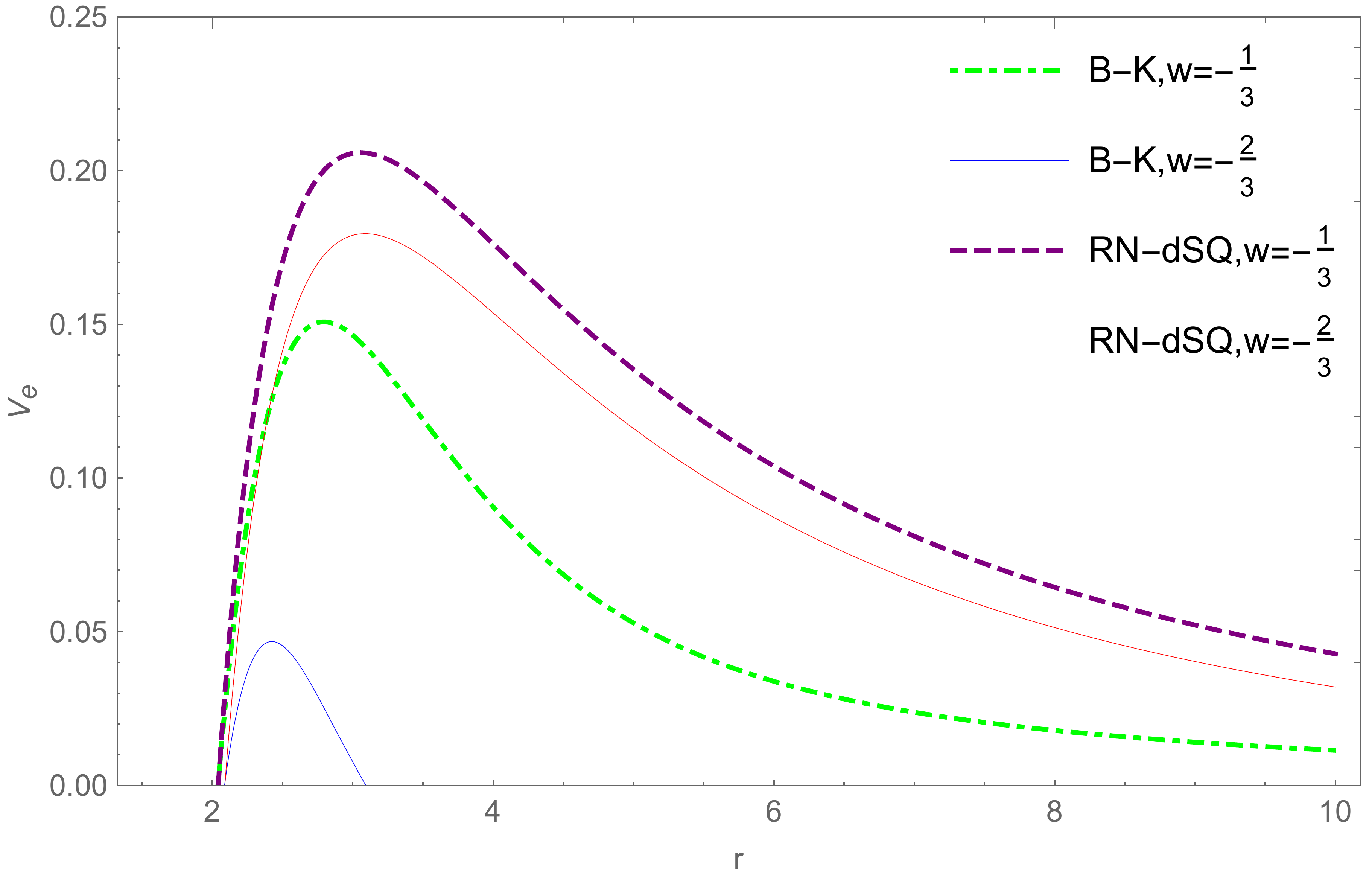}}
\centerline{(c)}
\end{minipage}
\hspace{7mm}
\begin{minipage}[t]{0.45\linewidth}
\centerline{\includegraphics[width=7.0cm]{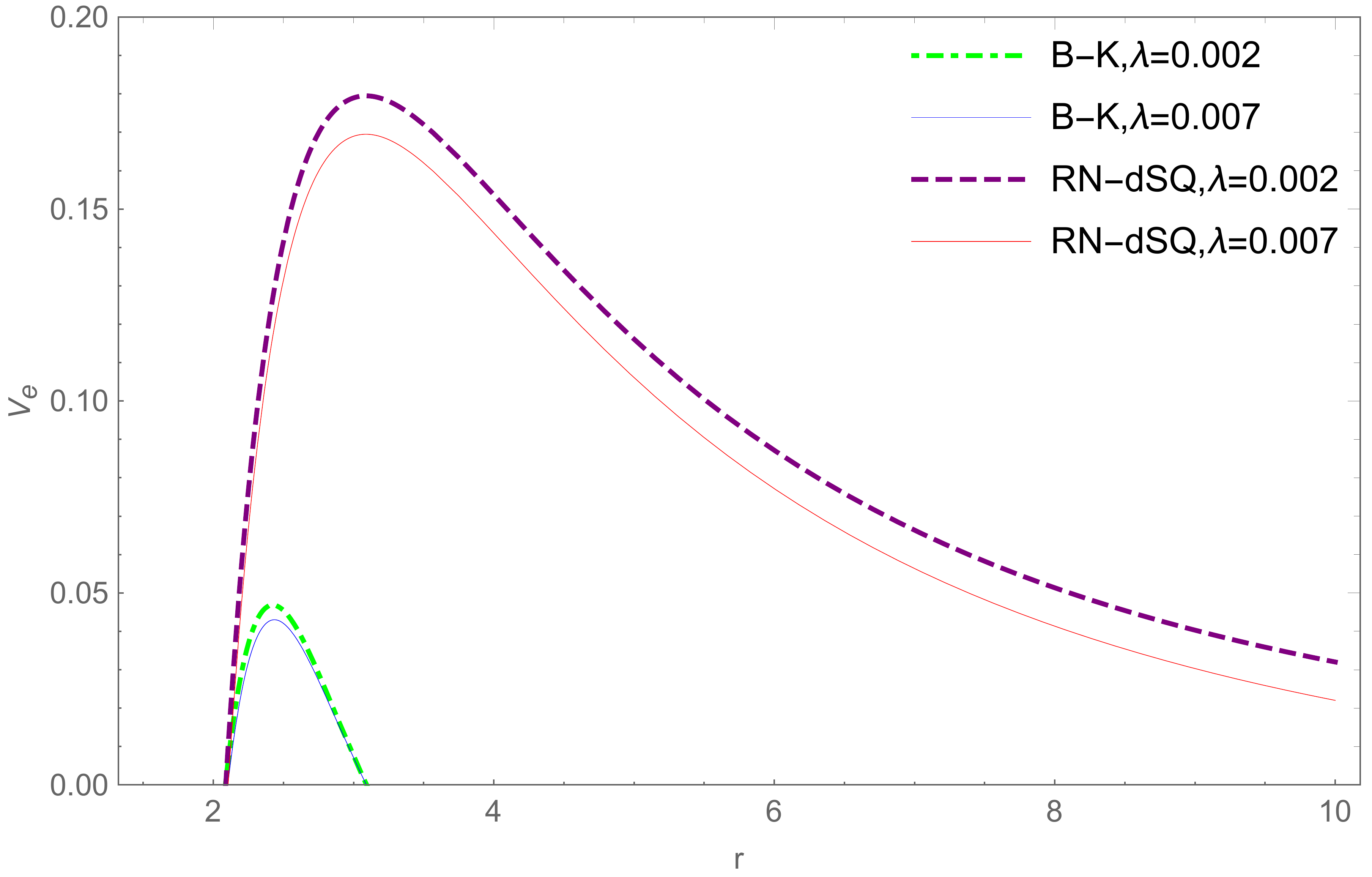}}
\centerline{(d)}
\end{minipage}
\end{tabular}
\caption{(a)The effective potential $V(r)$ for Bardeen-Kiselev BH with cosmological constant and RN-dSQ for $M=1$, $c=0.01$ $q=0.1$, $l=2$ and $\lambda=0.002$ in gravitational perturbations;
(b)The effective potential $V(r)$ for Bardeen-Kiselev BH with cosmological constant and RN-dSQ for $M=1$, $c=0.01$ $q=0.2$, $l=2$ and $w=-2/3$ in gravitational perturbations;(c)The effective potential $V(r)$ for Bardeen-Kiselev BH with cosmological constant and RN-dSQ for $M=1$, $c=0.01$ $q=0.1$, $l=2$ and $\lambda=0.002$ in electromagnetic perturbations;(d)The effective potential $V(r)$ for Bardeen-Kiselev BH with cosmological constant and RN-dSQ for $M=1$, $c=0.01$ $q=0.1$, $l=2$ and $w=-2/3$ in electromagnetic perturbations.}
\label{fig1}
\end{figure}

Moreover, since the solution of RN-dSQ is deduced from linear electromagnetic theory, $L(F)$ linearly depends on field strength $F$ which implies that $\widetilde{L}^{\prime}_{\tilde{F}}=\widetilde{L}_{\tilde{F} \tilde{F}}=0$. For the gravitational perturbations and electromagnetic perturbations, we present the effective potentials $V(r)$ for Bardeen-Kiselev BH with cosmological constant and the RN-dSQ in Fig.\ref{fig1}. Specifically speaking, the overall nature of both potentials for Bardeen-Kiselev BH with cosmological constant and the RN-dSQ are the same, i.e. they are positive definite between the event and cosmological horizons, and they all have a single maxima. Besides, for a fixed set of parameters in electromagnetic perturbation, the height of RN-dSQ potential is larger than the Bardeen-Kiselev one which implies that the RN-dSQ has smaller absorption coefficient than Bardeen-Kiselev BH with cosmological constant. Meanwhile, the opposing observation appears in gravitational perturbation. Through the figure, it shows that the potential decreases with increasing parameters $\lvert w \lvert$ or $\lambda$, which indicates that the smaller value of $\lvert w \lvert$ or $\lambda$ suppresses the emission modes for gravitational perturbation and electromagnetic perturbation.

{\centering  \section{The QNMs of the Bardeen-Kiselev black hole with cosmological constant for gravitational perturbations and electromagnetic perturbations } \label{secIV} }

In this section, our purpose is to discuss the QNMs and the stability of the perturbations in Bardeen-Kiselev BH with cosmological constant spacetime, especially for the QNMs of the Bardeen-Kiselev BH with cosmological constant for gravitational perturbations and electromagnetic perturbations. QNMs for a perturbed BH spacetime are the solutions to the wave equation given in Eq.\eqref{eq15} and Eq.\eqref{eq25}, and in order to derive these solutions, it is necessary to impose proper boundary conditions:

1. pure ingoing waves at the event horizon $\Psi(r) \sim e^{- i{\omega}r_{*}}$, $r_{*}$ $\rightarrow$  $-\infty$ $(r \rightarrow r_{+})$,

2. pure outgoing waves at the spatial infinity $\Psi(r) \sim e^{ i{\omega}r_{*}}$, $r_{*}$ $\rightarrow$  $\infty$ $(r \rightarrow r_{c})$.

However, it is tricky to analytically solve the time-independent, second-order differential equation (such as Eq.\eqref{eq15}) with the potential (such as Eq.\eqref{eq14}) for a nonlinear magnetic charged BH with cosmological constant surrounded by quintessence. The WKB method can be used for the effective potential, which configurates the form of a potential barrier and takes constant values at the event horizon and spatial infinity. Specifically speaking, this method  is based on matching the asymptotic WKB solutions at spatial infinity and the event horizon with a Taylor expansion near the top of the potential barrier through two turning points. By using the considered potential functions, one can obtain the QNMs frequencies through a sixth-order WKB method, and as seen in Ref.\cite{ret32}, this method is the most accurate one for finding the quasinormal spectrum with lower overtones.
The BH potential $V(r)$ in the present of the sixth order formula is
\begin{equation}
\frac{i(\omega^{2}-V_{0})}{\sqrt{-2V_{0}^{''}}}-\sum_{i=2}^{6}\varLambda_{i}=n+\frac{1}{2}, n=0,1,2,..., \label{eq27}    \end{equation}
among them, $V_{0}$ is the maximum effective potential of $V(r)$ at the tortoise coordinate $r_{*}$, $n$ is the overtone number (we study the case $n=0$) and the correction term $\varLambda_{i}$ can be obtained in Ref.\cite{ret20}. And generally speaking, the quasinormal frequencies $\omega$ take the form $\omega=\omega_{R}-i\omega_{I}$, where the real part and the imaginary part of $\omega$ denote actual field oscillation and damping of the perturbation respectively.

\begin{figure}[htbp]
\begin{tabular}{cc}
\begin{minipage}[t]{0.45\linewidth}
\centerline{\includegraphics[width=7.0cm]{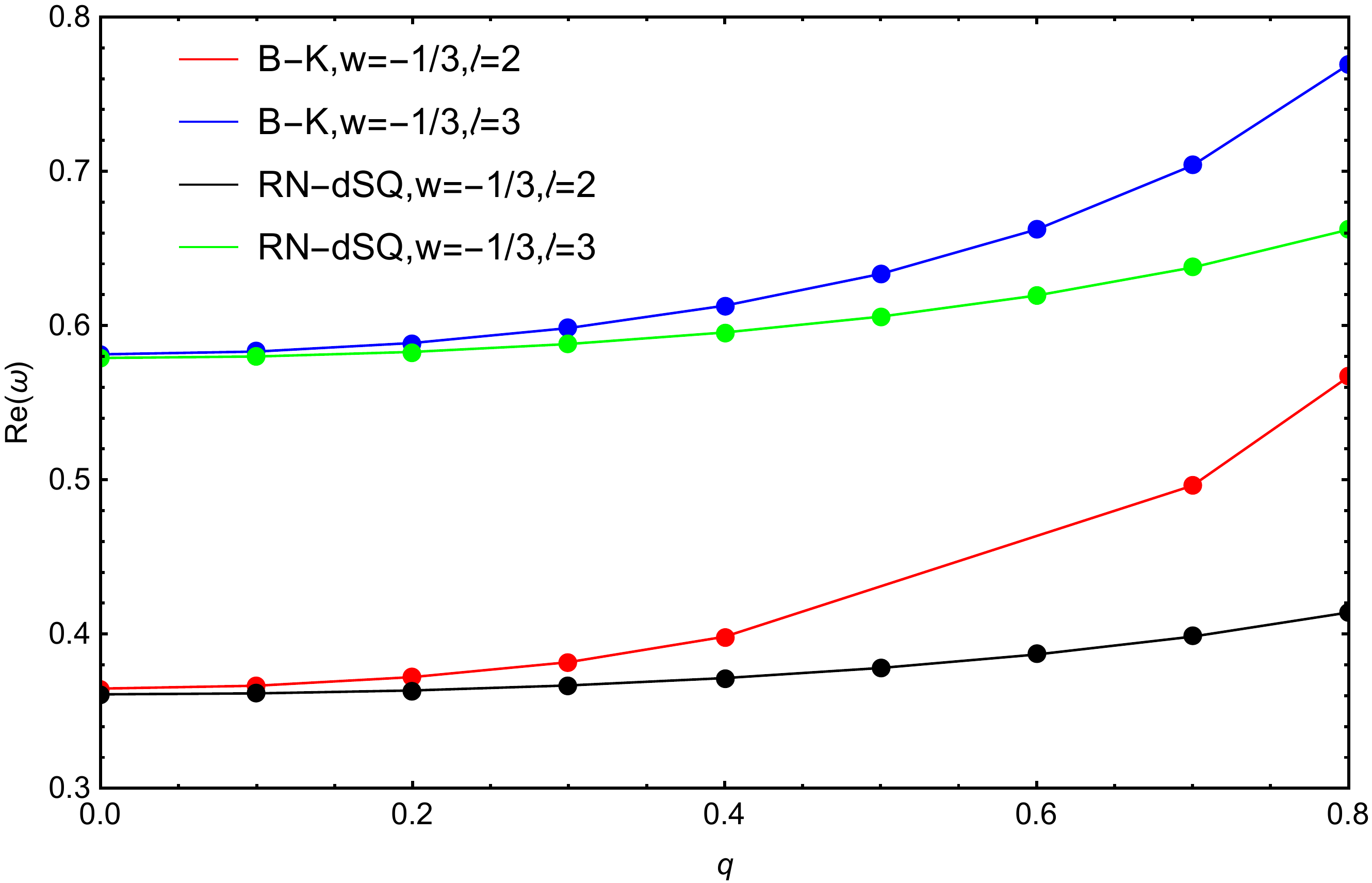}}
\centerline{(a)}
\end{minipage}
\hspace{7mm}
\begin{minipage}[t]{0.45\linewidth}
\centerline{\includegraphics[width=7.0cm]{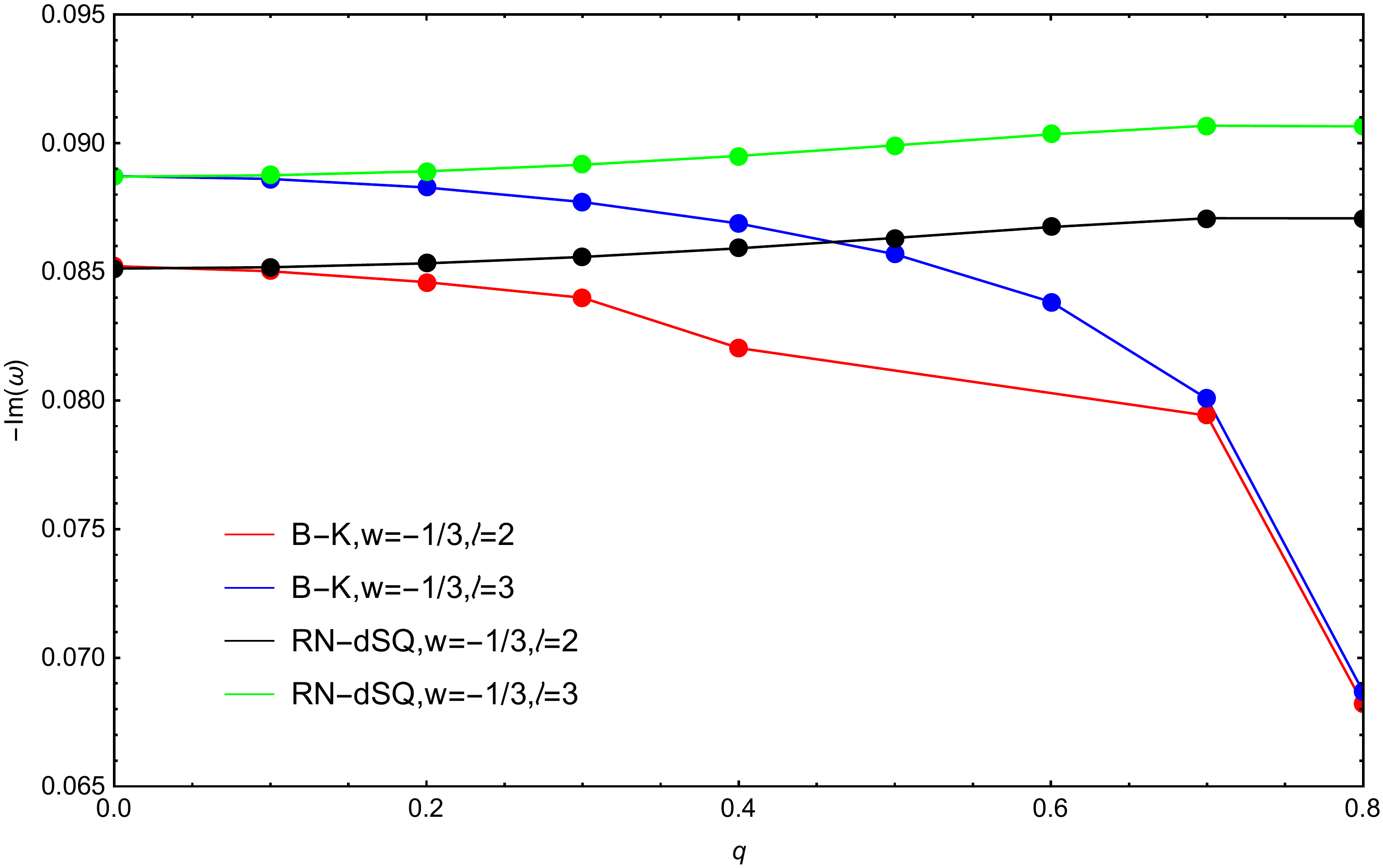}}
\centerline{(b)}
\end{minipage}
\\
\begin{minipage}[t]{0.45\linewidth}
\centerline{\includegraphics[width=7.0cm]{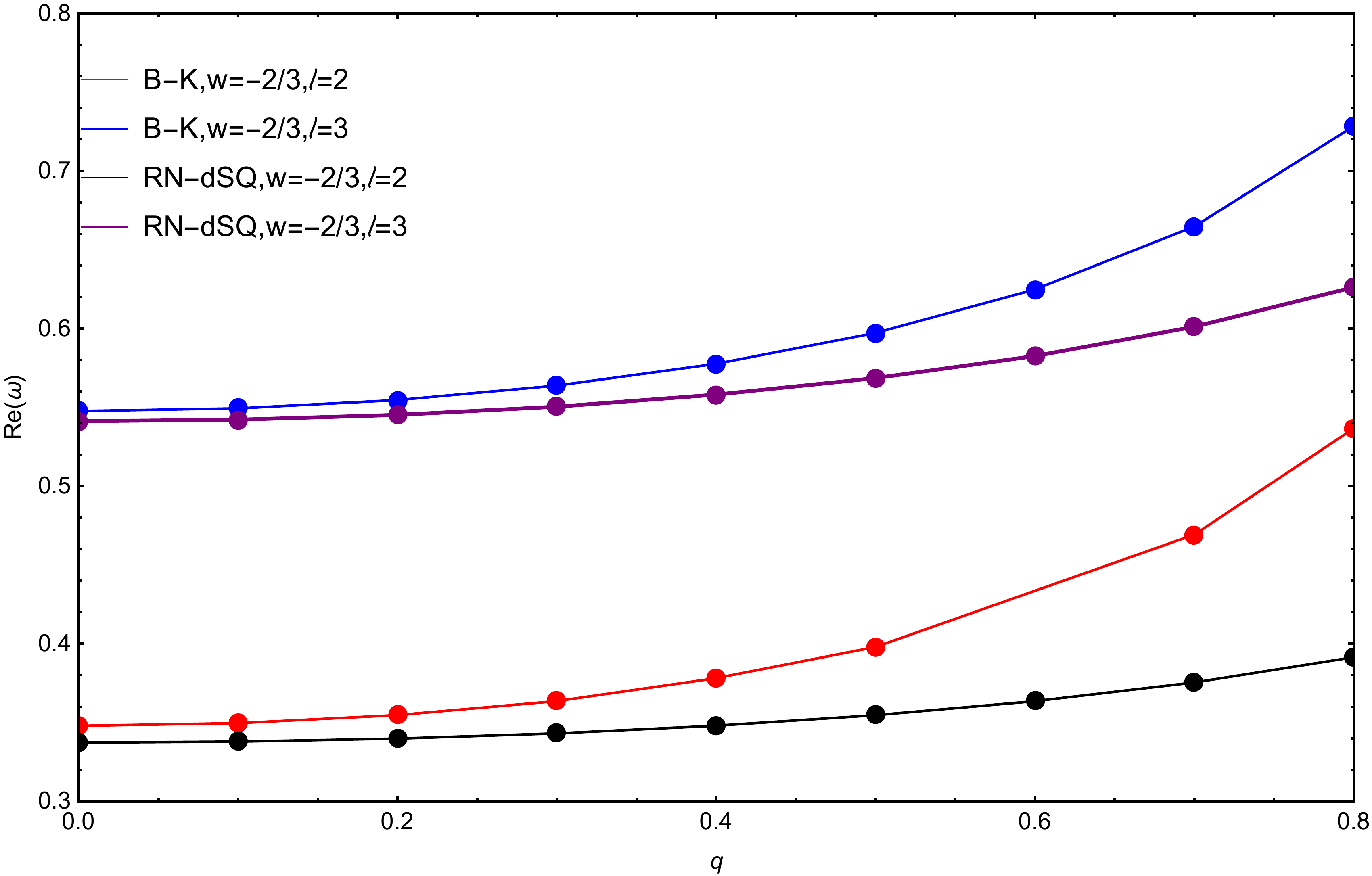}}
\centerline{(c)}
\end{minipage}
\hspace{7mm}
\begin{minipage}[t]{0.45\linewidth}
\centerline{\includegraphics[width=7.0cm]{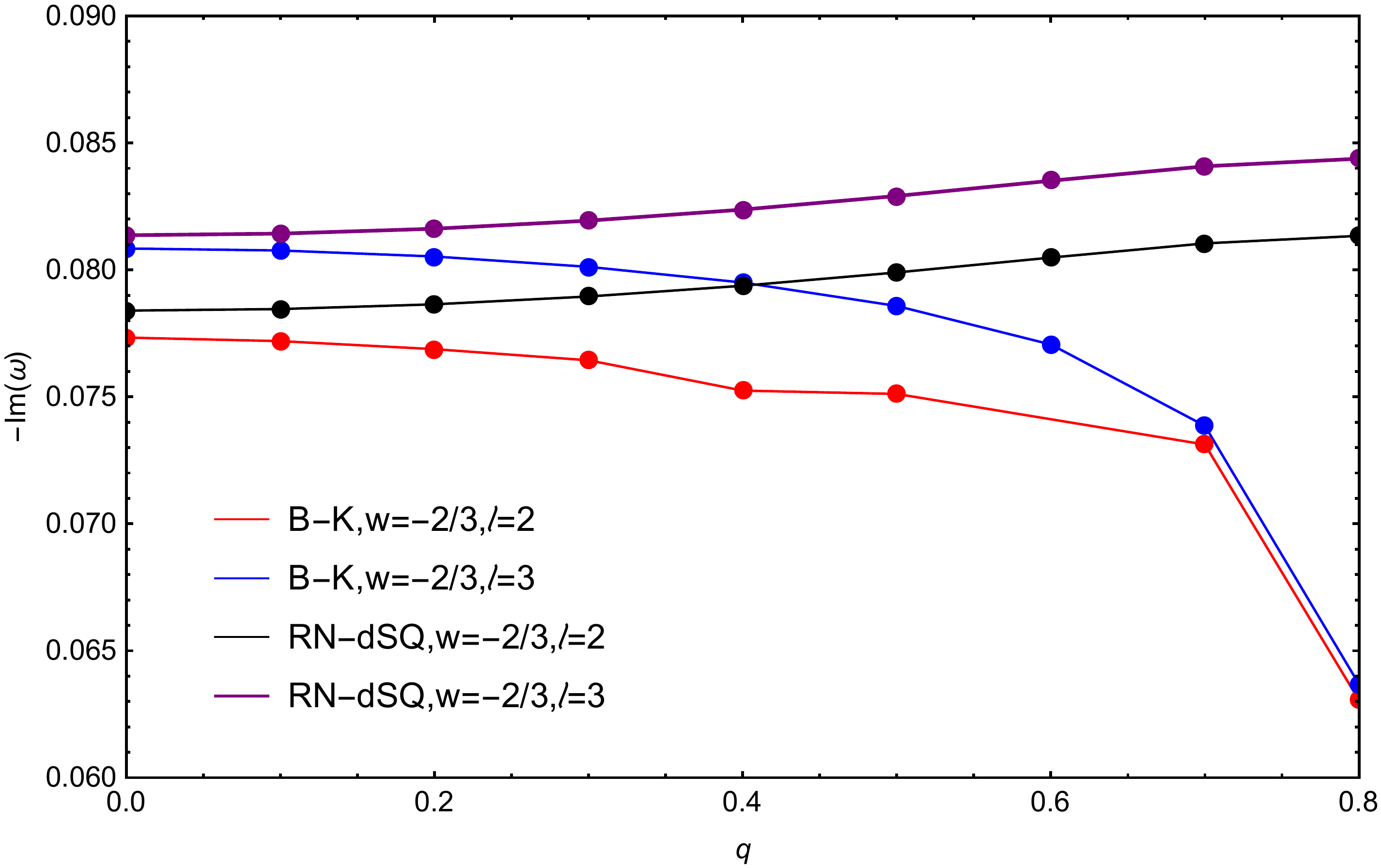}}
\centerline{(d)}
\end{minipage}
\end{tabular}
\caption{(a)Variation of Re $\omega$ with the magnetic charge $q$ for the state parameter $w=-1/3$; (b) Variation of -Im $\omega$ with the magnetic charge $q$ for the state parameter $w=-1/3$; (c)Variation of Re $\omega$ with the magnetic charge $q$ for the state parameter $w=-2/3$; (d) Variation of -Im $\omega$ with the magnetic charge $q$ for the state parameter $w=-2/3$. In both cases we take $M=1$, $c=0.01$ and $\lambda=0.001$. }
\label{fig2}
\end{figure}

\begin{figure}[htbp]
\begin{tabular}{cc}
\begin{minipage}[t]{0.45\linewidth}
\centerline{\includegraphics[width=7.0cm]{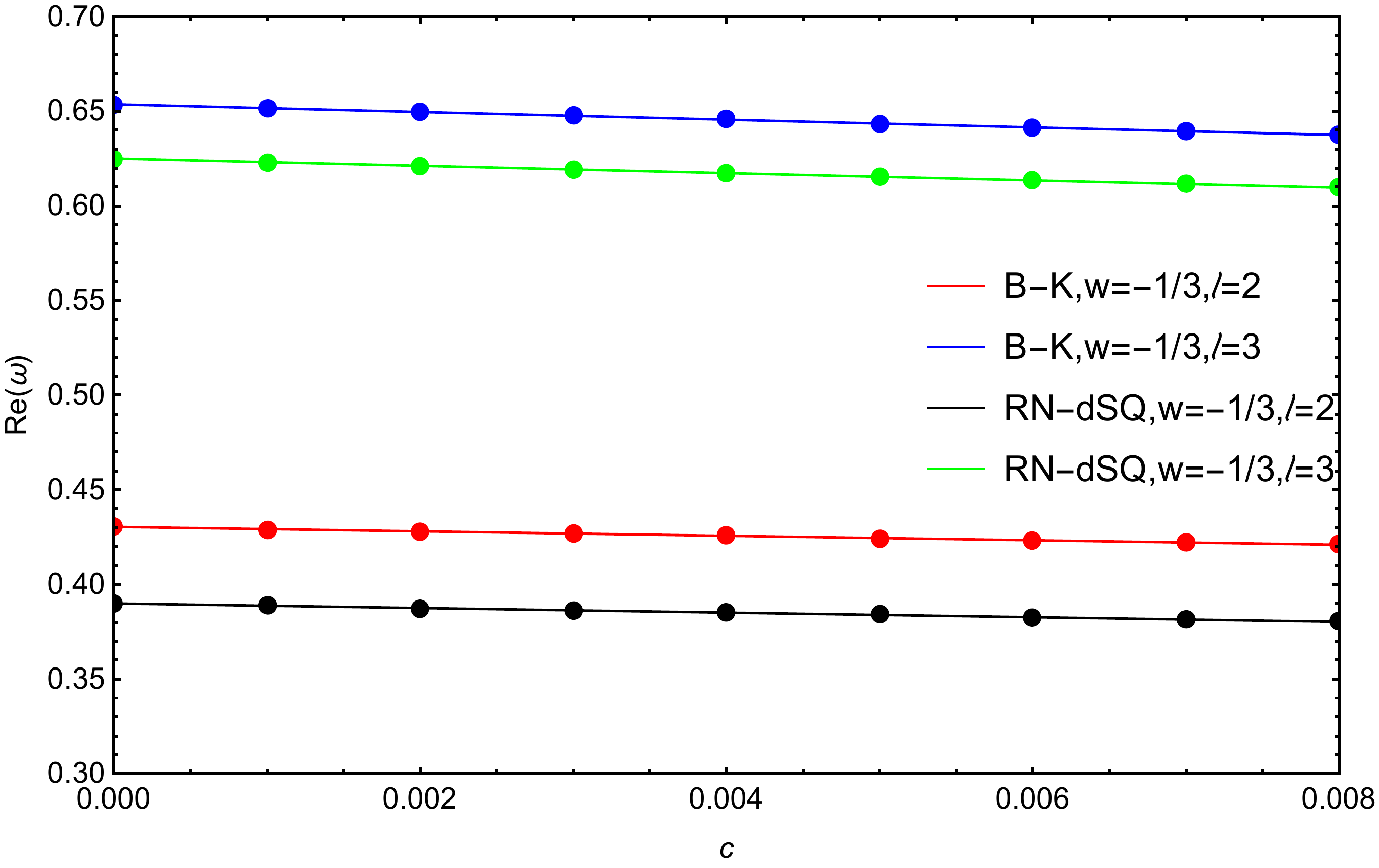}}
\centerline{(a)}
\end{minipage}
\hspace{7mm}
\begin{minipage}[t]{0.45\linewidth}
\centerline{\includegraphics[width=7.0cm]{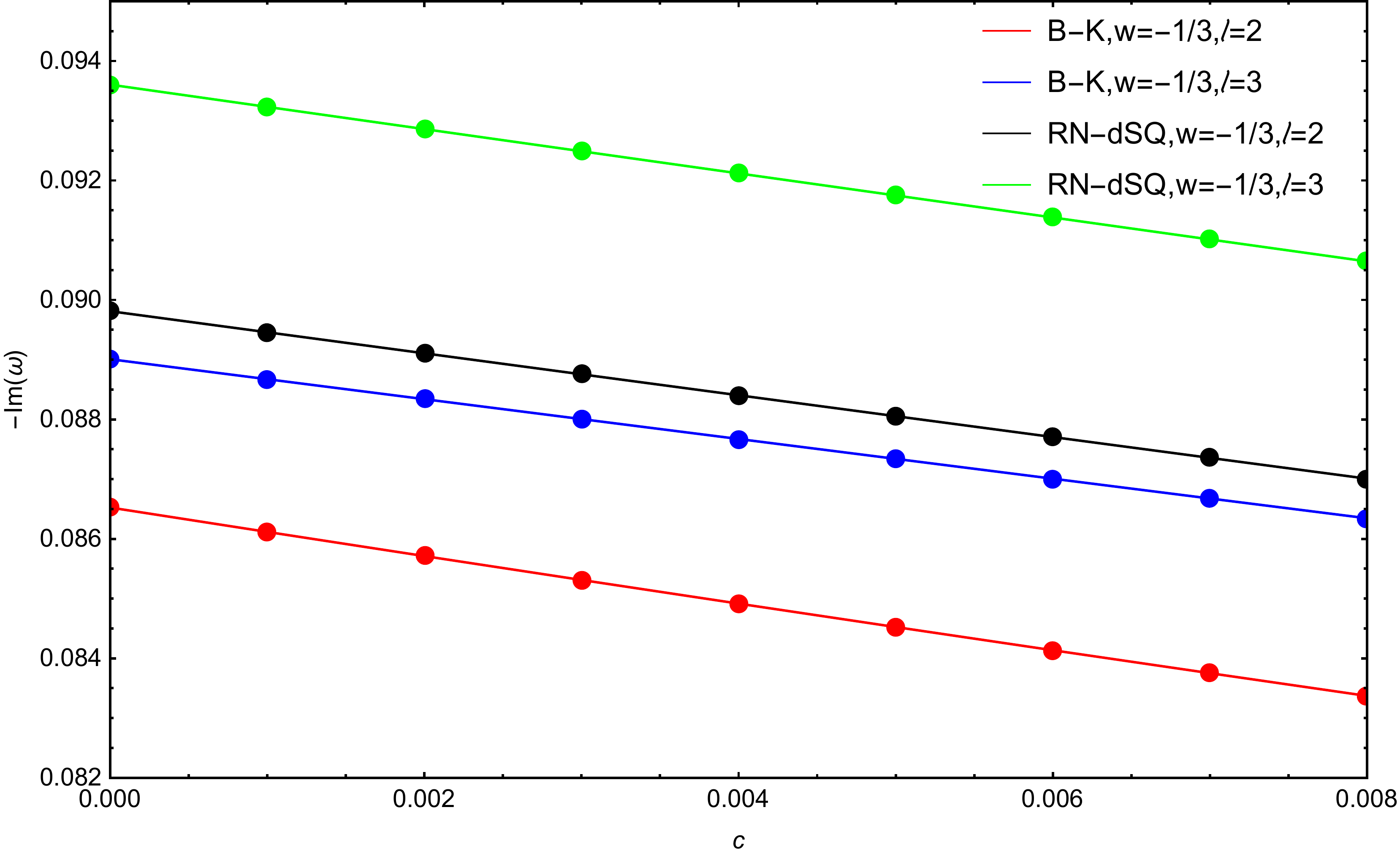}}
\centerline{(b)}
\end{minipage}
\\
\begin{minipage}[t]{0.45\linewidth}
\centerline{\includegraphics[width=7.0cm]{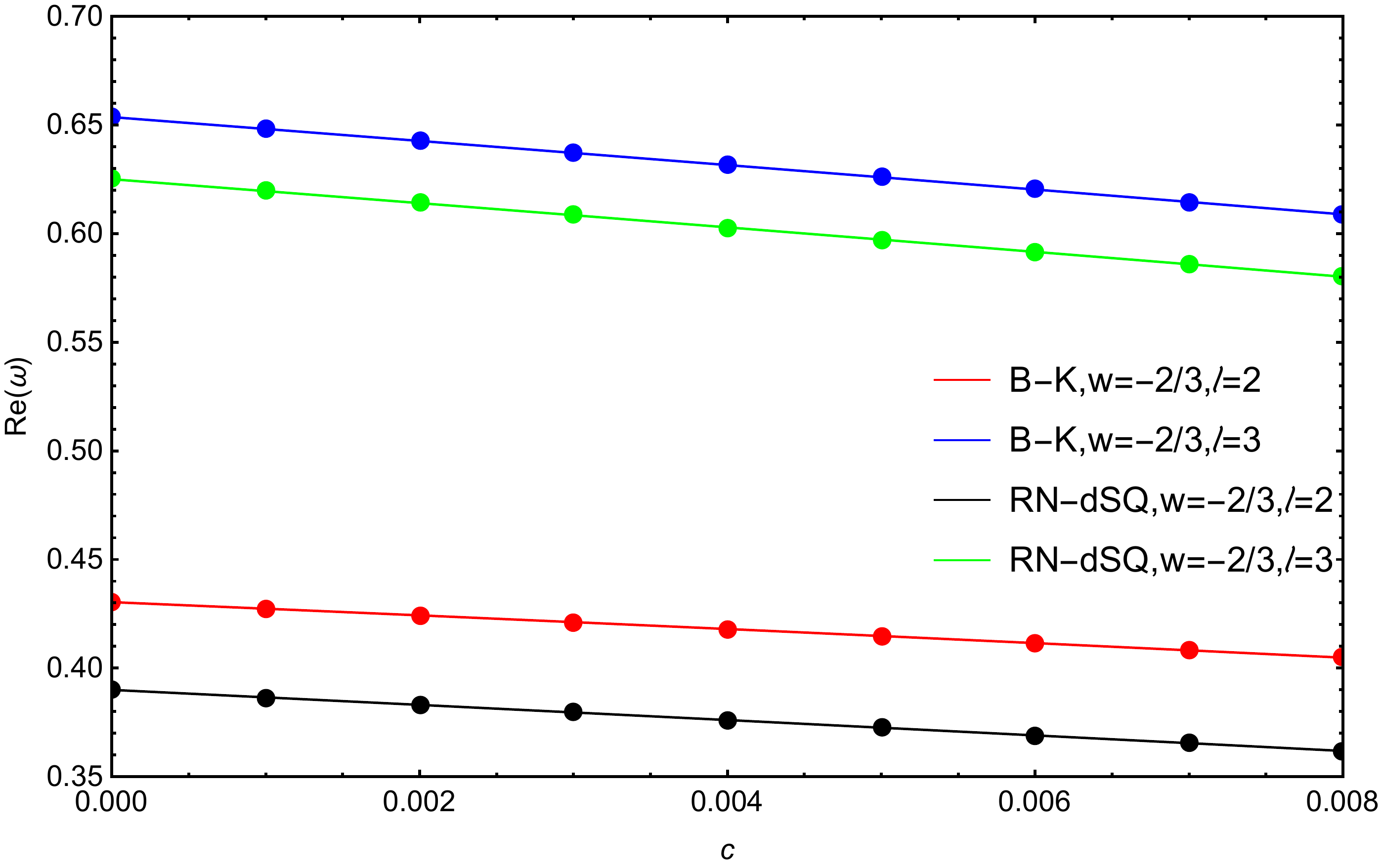}}
\centerline{(c)}
\end{minipage}
\hspace{7mm}
\begin{minipage}[t]{0.45\linewidth}
\centerline{\includegraphics[width=7.0cm]{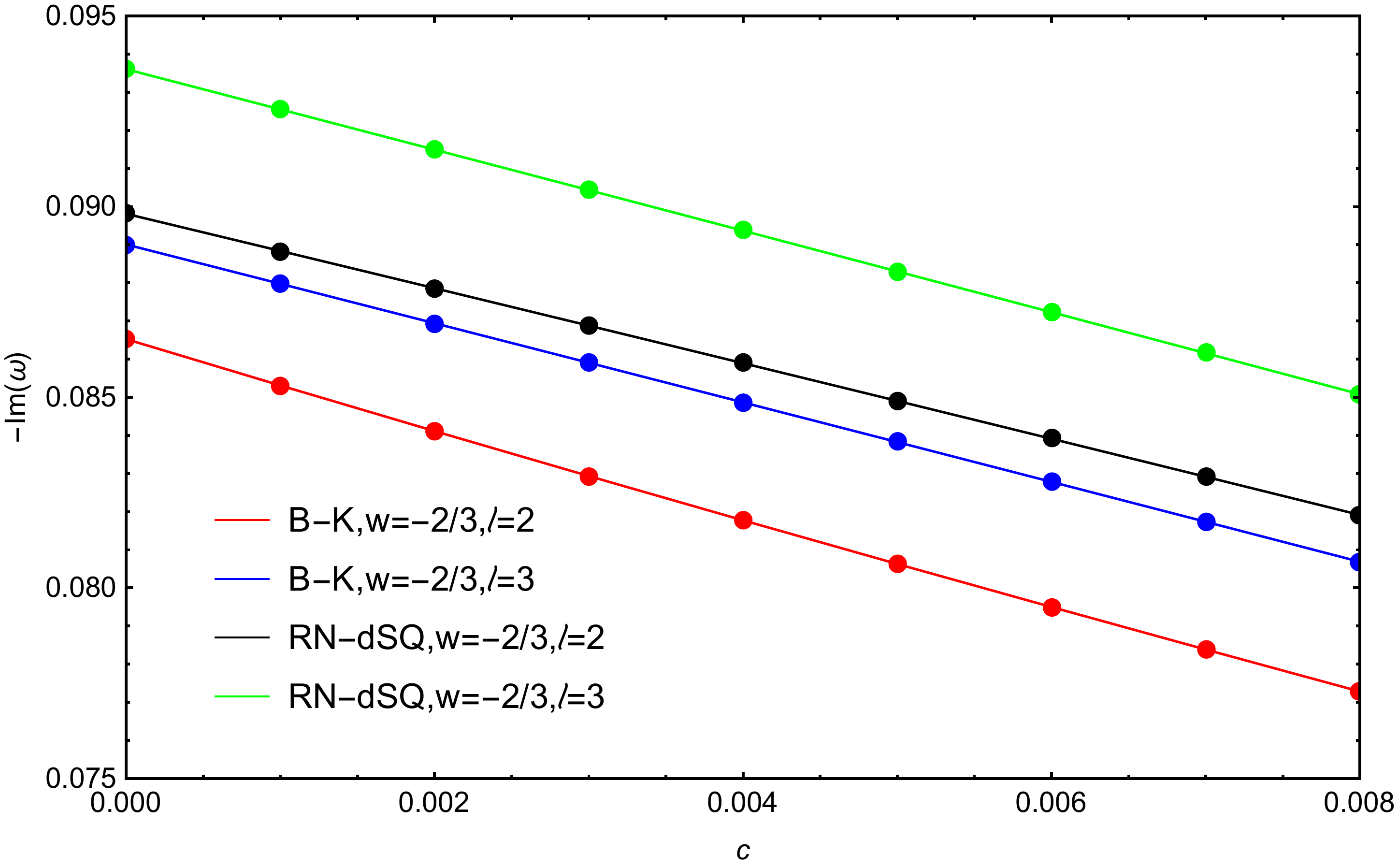}}
\centerline{(d)}
\end{minipage}
\end{tabular}
\caption{(a)Variation of Re $\omega$ with the normalization factor $c$ for the state parameter $w=-1/3$; (b) Variation of -Im $\omega$ with the normalization factor $c$ for the state parameter $w=-1/3$; (c)Variation of Re $\omega$ with the normalization factor $c$ for the state parameter $w=-2/3$; (d) Variation of -Im $\omega$ with the normalization factor $c$ for the state parameter $w=-2/3$. In both cases we take $M=1$, $q=0.5$ and $\lambda=0.001$.}
\label{fig3}
\end{figure}

\begin{figure}[htbp]
\begin{tabular}{cc}
\begin{minipage}[t]{0.45\linewidth}
\centerline{\includegraphics[width=7.0cm]{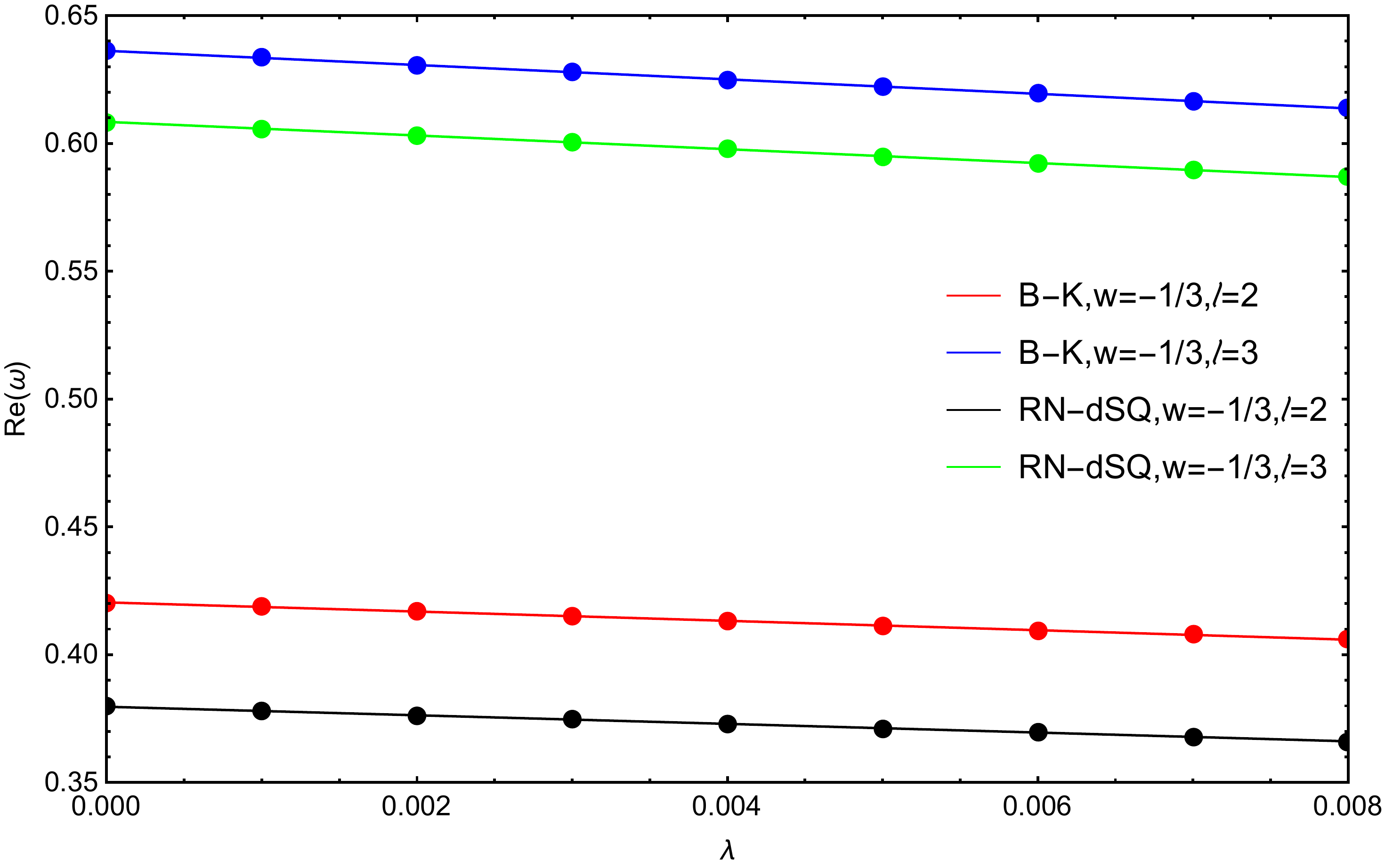}}
\centerline{(a)}
\end{minipage}
\hspace{7mm}
\begin{minipage}[t]{0.45\linewidth}
\centerline{\includegraphics[width=7.0cm]{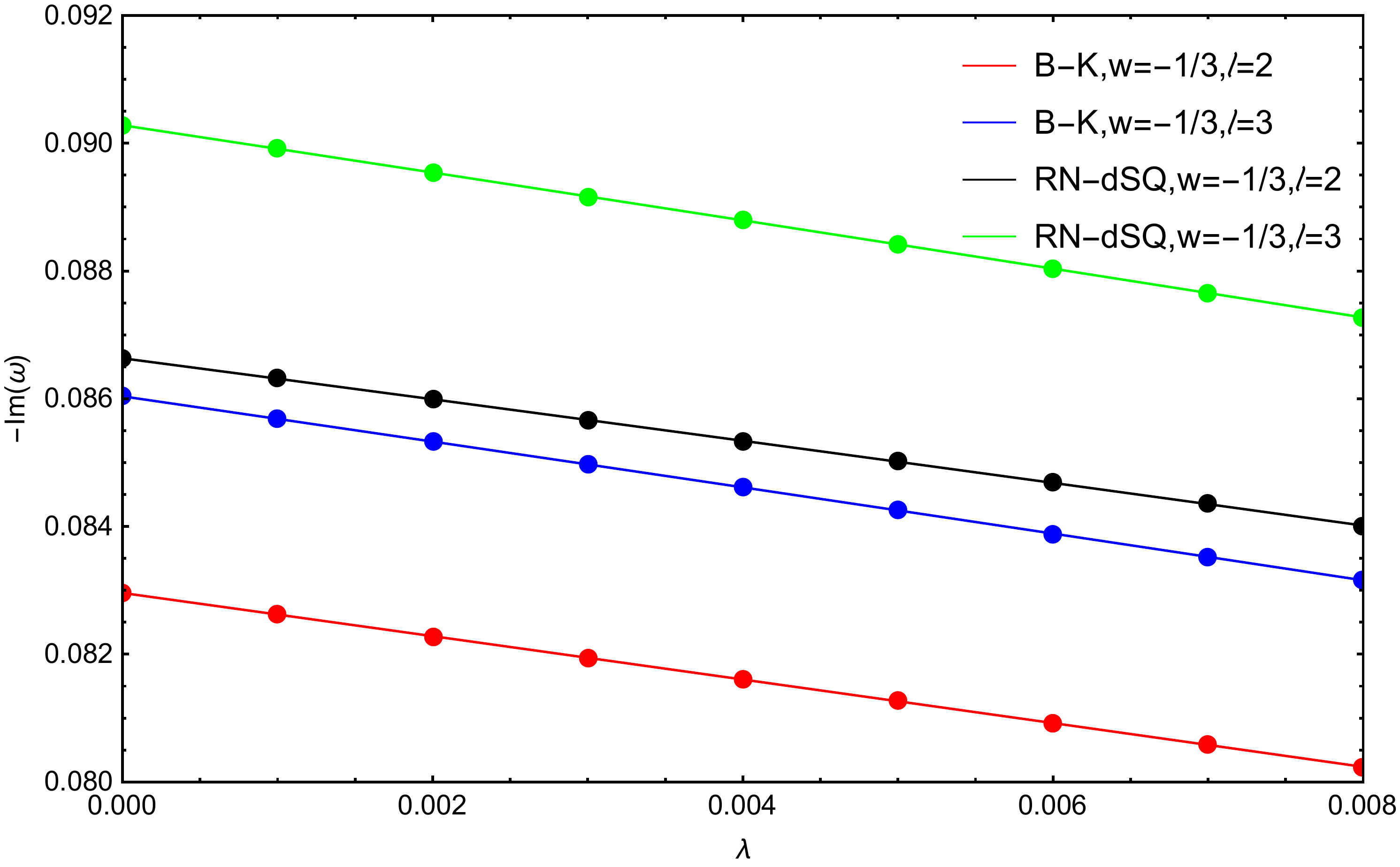}}
\centerline{(b)}
\end{minipage}
\\
\begin{minipage}[t]{0.45\linewidth}
\centerline{\includegraphics[width=7.0cm]{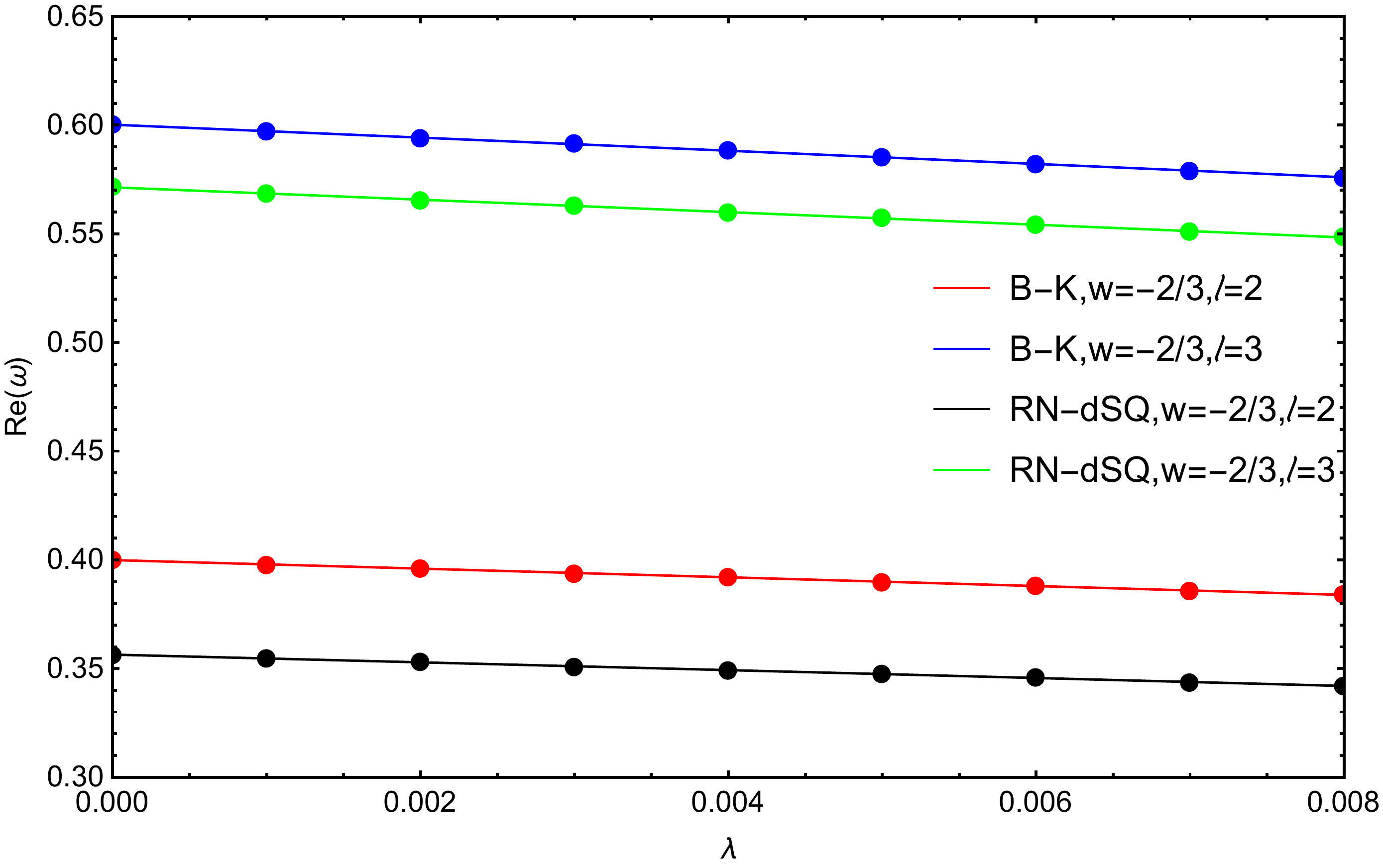}}
\centerline{(c)}
\end{minipage}
\hspace{7mm}
\begin{minipage}[t]{0.45\linewidth}
\centerline{\includegraphics[width=7.0cm]{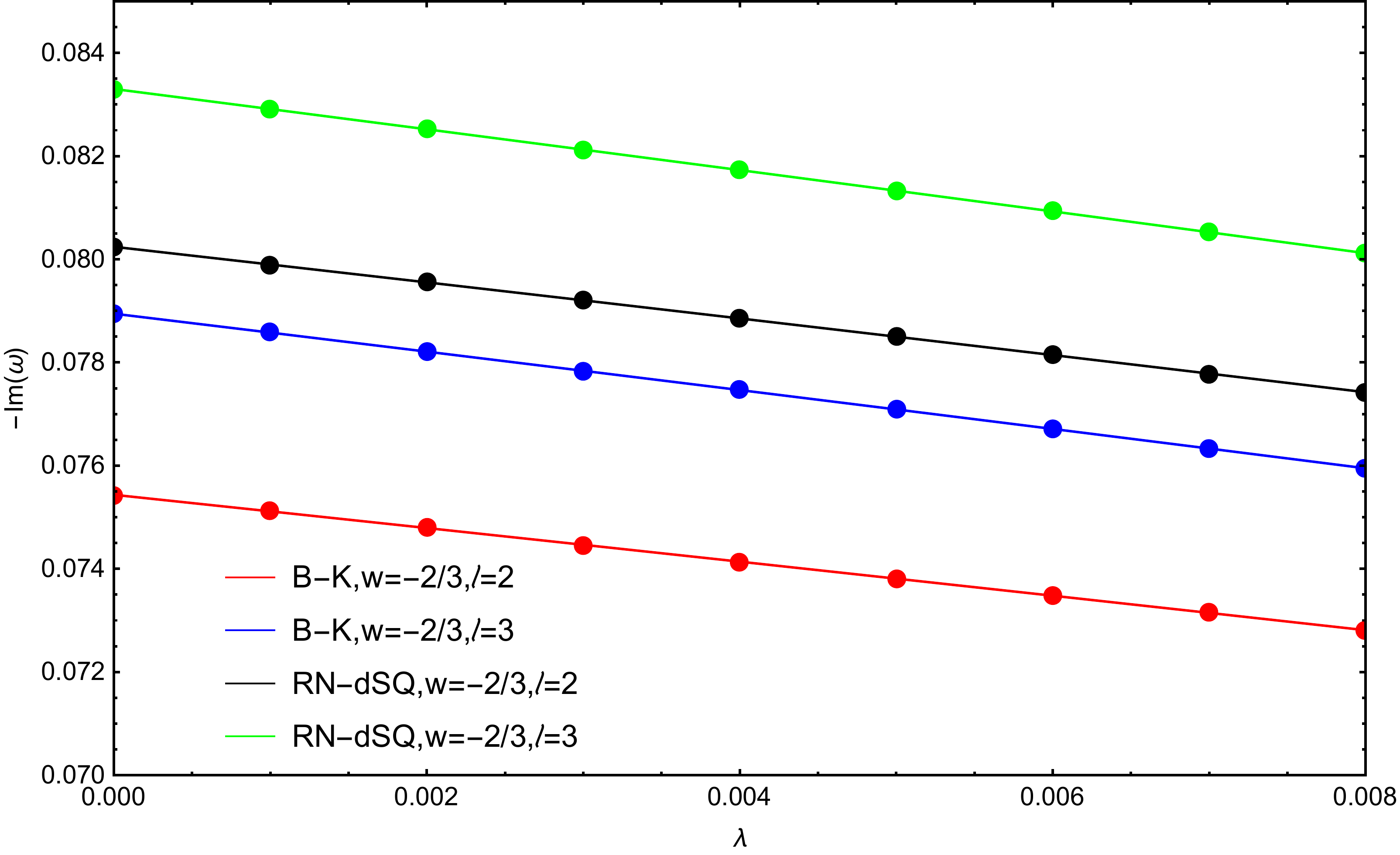}}
\centerline{(d)}
\end{minipage}
\end{tabular}
\caption{(a)Variation of Re $\omega$ with the cosmological constant $\lambda$ for the state parameter $w=-1/3$; (b) Variation of -Im $\omega$ with the cosmological constant $\lambda$ for the state parameter $w=-1/3$; (c)Variation of Re $\omega$ with the cosmological constant $\lambda$ for the state parameter $w=-2/3$; (d) Variation of -Im $\omega$ with the cosmological constant $\lambda$ for the state parameter $w=-2/3$. In both cases we take $M=1$, $q=0.5$ and $c=0.01$.}
\label{fig4}
\end{figure}

\begin{table}[htbp]
\caption{QNM frequencies of electromagnetic field perturbation for Bardeen-Kiselev BH with cosmological constant and RN-dSQ.}
\label{tab1}
\begin{tabular}{lllll}
\hline
\multicolumn{5}{l}{M=1, c=0.01, $\lambda=0.001$,  w=-1/3}
  \\ \hline
\multicolumn{2}{c}{}             & $\ell=2$                    &             &  $\ell=3$
 \\ \hline
\multicolumn{1}{c}{} &RN-dSQ                            &Bardeen-Kiselev                        & RN-dSQ                   & Bardeen-Kiselev
\\  \hline
q=0.1         & 0.442872 -0.0908932 i           & 0.36724   -0.119367 i           & 0.635522 -0.0914572 i         & 0.535244   -0.121049 i         \\
q=0.2         & 0.445194 -0.0910523 i          & 0.495555   -0.112316 i          & 0.638784 -0.0916127 i           & 0.693571   -0.110998 i         \\
q=0.3         & 0.449182 -0.0913118 i           & 0.53953   -0.104102 i           & 0.644385 -0.0918663 i         & 0.744857   -0.10253 i          \\
q=0.4         & 0.455032 -0.0916603 i         & 0.557533   -0.0977495 i         & 0.652595 -0.092206 i        & 0.76412 -0.0965164 i           \\
q=0.5         & 0.463062 -0.0920738 i          & 0.564534   -0.0929304 i         & 0.663858 -0.0926075 i        & 0.770024   -0.0921494 i        \\
q=0.6         & 0.473784 -0.0925008 i           & 0.565716   -0.089275 i          & 0.678878 -0.0930191  i        & 0.771675   -0.0888982 i        \\
 \hline
\multicolumn{5}{l}{M=1, c=0.01, $\lambda=0.001$, w=-2/3}                                                                                                      \\ \hline
\multicolumn{2}{c}{}   & $\ell=2$                               &                 &$\ell=3$                                                                       \\ \hline
\multicolumn{1}{c}{} &RN-dSQ                   &Bardeen-Kiselev                      &RN-dSQ               &Bardeen-Kiselev                 \\ \hline
q=0.1         & 0.414383 -0.0831505 i             & 0.198681 -0.152195 i            & 0.594251 -0.0836431 i           & 0.306161 -0.14864 i            \\
q=0.2         & 0.416746 -0.0833475 i             & 0.369186 -0.129174 i            & 0.59758 -0.0838376 i             & 0.532669 -0.126129 i           \\
q=0.3         & 0.420804 -0.0836725 i             & 0.437717 -0.118541 i            & 0.603296 -0.0841582 i           & 0.619425 -0.115049 i           \\
q=0.4         & 0.426758 -0.0841178 i            & 0.471818 -0.110363 i            & 0.611676 -0.084597 i           & 0.660201 -0.10686 i           \\
q=0.5         & 0.434935 -0.0846656 i            & 0.489898 -0.103874 i            & 0.623176 -0.0851359 i           & 0.68026 -0.100461 i            \\
q=0.6         & 0.445858 -0.0852751 i           & 0.498972 -0.0987989 i           &0.638521 -0.0857336 i          & 0.688793 -0.0954421 i          \\
q=0.7         & 0.460387 -0.0858485 i            & 0.502352 -0.0949737 i           &0.658901 -0.0862925 i          & 0.689691 -0.091777 i           \\
 \hline
\end{tabular}
\end{table}

\begin{table}[htbp]
\caption{QNM frequencies of electromagnetic field perturbation for Bardeen-Kiselev BH with cosmological constant and RN-dSQ .}
\label{tab2}
\begin{tabular}{lllll}
\hline
\multicolumn{5}{l}{M=1, q=0.5, $\lambda=0.001$,  w=-1/3}
  \\ \hline
\multicolumn{2}{c}{}             & $\ell=2$                   &             & $\ell=3$
 \\ \hline
\multicolumn{1}{c}{} &RN-dSQ                            &Bardeen-Kiselev                        & RN-dSQ                    & Bardeen-Kiselev
\\  \hline
c=0     & 0.477653 -0.0958952 i & 0.60626   -0.0854858 i  & 0.684955 -0.0964615  i & 0.821065   -0.0871122 i    \\
c=0.001 & 0.476187 -0.0955096 i  & 0.602076   -0.086518 i  & 0.682835 -0.0960726  i   & 0.81586   -0.0878153 i         \\
c=0.002 & 0.474722 -0.0951248 i & 0.597886   -0.0874856 i &0.680717 -0.0956845 i   & 0.810674   -0.0884689 i          \\
c=0.003 & 0.473259 -0.0947407 i & 0.593691   -0.0883888 i & 0.678601 -0.0952971 i & 0.805508   -0.0890751 i          \\
c=0.004 & 0.471798 -0.0943574 i & 0.589495   -0.0892268 i & 0.676488 -0.0949106 i & 0.800365 -0.0896358 i           \\
c=0.005 &0.470338 -0.0939749 i & 0.585301   -0.0899992 i & 0.674377 -0.0945248 i & 0.795244   -0.0901527 i        \\
c=0.006 & 0.46888 -0.0935932 i & 0.581113   -0.0907068 i & 0.672268 -0.0941398 i & 0.790147   -0.0906279 i         \\
c=0.007 & 0.467423 -0.0932122 i & 0.576938   -0.0913509 i & 0.670162 -0.0937555 i & 0.785076   -0.0910633 i        \\
c=0.008 & 0.465968 -0.0928319 i & 0.57278   -0.0919342 i  & 0.668058 -0.0933721 i & 0.780031 -0.0914606 i        \\ \hline
\multicolumn{5}{l}{M=1, q=0.5, $\lambda=0.001$,  w=-2/3}                                                                                                      \\ \hline
\multicolumn{2}{c}{}   &  $\ell=2$                                &                 &$\ell=3$                                                                       \\ \hline
\multicolumn{1}{c}{} &RN-dSQ                  &Bardeen-Kiselev                      &RN-dSQ               &Bardeen-Kiselev                 \\ \hline
c=0     & 0.477653 -0.0958952 i   & 0.60626 -0.0854858 i    & 0.684955 -0.0964615 i   & 0.821065 -0.0871122 i                \\
c=0.001 & 0.473505 -0.0947854 i   & 0.595604 -0.0916735 i   & 0.678952 -0.095342 i   & 0.806694 -0.0909355 i             \\
c=0.002 & 0.469331 -0.0936729 i   & 0.582242 -0.0972697 i   & 0.672913 -0.0942198 i   & 0.791951 -0.0941978 i           \\
c=0.003 &  0.465131 -0.0925576 i   & 0.568667 -0.100549 i    & 0.666837 -0.0930948i   & 0.777026 -0.0966223 i           \\
c=0.004 &  0.460904 -0.0914394i    & 0.556067 -0.102324 i    & 0.660722 -0.091967 i   & 0.762317 -0.0982896 i          \\
c=0.005 &  0.456649 -0.0903182i    & 0.544186 -0.103322 i    & 0.654568 -0.0908362 i   & 0.747943 -0.0993956 i           \\
c=0.006 &  0.452366 -0.089194i    & 0.532764 -0.103887 i    & 0.648375 -0.0897024 i    & 0.733896 -0.100096 i          \\
c=0.007 &  0.448054 -0.0880668 i    & 0.521673 -0.104168 i    & 0.64214 -0.0885656  i    & 0.720137 -0.100496 i             \\
c=0.008 & 0.443712 -0.0869363 i    & 0.510851 -0.104236 i    & 0.635862 -0.0874256 i     & 0.70663 -0.100662 i          \\ \hline
\end{tabular}
\end{table}

\begin{table}[htbp]
\caption{QNM frequencies of electromagnetic field perturbation for Bardeen-Kiselev BH with cosmological constant and RN-dSQ .}
\label{tab3}
\begin{tabular}{lllll}
\hline
\multicolumn{5}{l}{M=1, q=0.5, c=0.01, w=-1/3}
  \\ \hline
\multicolumn{2}{c}{}             & $\ell=2$                   &             & $\ell=3$
 \\ \hline
\multicolumn{1}{c}{} &RN-dSQ                            &Bardeen-Kiselev                       & RN-dSQ                   &Bardeen-Kiselev
\\  \hline
$\lambda=0$      & 0.465071 -0.0924752 i  & 0.567498   -0.0933613 i & 0.666761 -0.0930134 i & 0.773717   -0.0925742 i   \\
$\lambda=0.001$  & 0.463062 -0.0920738 i  & 0.564534   -0.0929304 i & 0.663858 -0.0926075 i & 0.770024   -0.0921494 i          \\
$\lambda=0.002$  & 0.461044 -0.0916706  i  & 0.561566   -0.0924977 i & 0.660941 -0.0921997 i & 0.76632   -0.0917227 i          \\
$\lambda=0.003$  & 0.459016 -0.0912657 i & 0.558593   -0.0920629 i & 0.658011 -0.0917902 i & 0.762605   -0.0912941 i          \\
$\lambda=0.004$  & 0.456979 -0.0908589 i & 0.555615 -0.0916263 i   & 0.655067 -0.0913788 i & 0.758877   -0.0908635 i           \\
$\lambda=0.005$  & 0.454931 -0.0904504 i & 0.552631 -0.0911876 i   & 0.65211 -0.0909656 i & 0.755138   -0.090431 i       \\
$\lambda=0.006$  & 0.452875 -0.09004 i & 0.549642   -0.0907469 i & 0.649139 -0.0905504 i  & 0.751386   -0.0899965 i          \\
$\lambda=0.007$  &0.450808 -0.0896277  i & 0.546647   -0.0903042 i & 0.646154 -0.0901334 i & 0.747622   -0.08956 i        \\
$\lambda=0.008$  & 0.448731 -0.0892135 i & 0.543646   -0.0898594 i & 0.643155 -0.0897144  i & 0.743845   -0.0891215 i        \\ \hline
\multicolumn{5}{l}{M=1, q=0.5, c=0.01, w=-2/3}                                                                                                      \\ \hline
\multicolumn{2}{c}{}   &  $\ell=2$                                &                 &$\ell=3$                                                                       \\ \hline
\multicolumn{1}{c}{} &RN-dSQ                      &Bardeen-Kiselev          &RN-dSQ               &Bardeen-Kiselev                \\ \hline
$\lambda=0$       & 0.437079 -0.085085  i    & 0.492648 -0.104614 i    & 0.626273 -0.0855598 i    & 0.683868 -0.101142 i              \\
$\lambda=0.001$  & 0.434935 -0.0846656 i    & 0.489898 -0.103874 i    & 0.623176 -0.0851359 i    & 0.68026 -0.100461 i              \\
$\lambda=0.002$  & 0.432779 -0.0842442 i    & 0.487139 -0.103133 i    & 0.620064 -0.0847098 i    & 0.676636 -0.0997798 i            \\
$\lambda=0.003$  & 0.430611 -0.0838206 i     & 0.484371 -0.102392 i    & 0.616935 -0.0842816 i   & 0.672998 -0.0990974 i          \\
$\lambda=0.004$  & 0.428432 -0.0833949 i    & 0.481594 -0.10165 i     & 0.61379 -0.0838512  i   & 0.669344 -0.0984141 i         \\
$\lambda=0.005$  & 0.426241 -0.082967  i    & 0.478808 -0.100907 i    & 0.610628 -0.0834186 i    & 0.665674 -0.0977298 i           \\
$\lambda=0.006$  & 0.424039 -0.0825368 i    & 0.476012 -0.100164 i    & 0.607449 -0.0829837 i   & 0.661989 -0.0970446 i          \\
$\lambda=0.007$  & 0.421824 -0.0821044 i    & 0.473206 -0.0994197 i   & 0.604254 -0.0825465  i   & 0.658287 -0.0963583 i           \\
$\lambda=0.008$  & 0.419596 -0.0816698 i   & 0.470391 -0.0986748 i   & 0.60104 -0.082107 i      & 0.654569 -0.095671 i         \\ \hline
\end{tabular}
\end{table}

We have numerically obtained the QNMs frequencies for the gravitational and electromagnetic perturbations. In Fig.\ref{fig2}, Fig.\ref{fig3} and Fig.\ref{fig4}, we have exploited the 6rd order WKB approximation for calculating the frequencies of the gravitational perturbations of Bardeen-Kiselev BH with cosmological constant and RN-dSQ, these frequencies can be obtained by individually varying $q$, $c$ and $\lambda$. In Fig.\ref{fig2}, for the same parameter space, one can see that the oscillation frequency of Bardeen-Kiselev BH with cosmological constant and RN-dSQ increase with the increasing magnetic charge $q$, for the Bardeen-Kiselev BH with cosmological constant, the damping rate decreases as the charge increases, which implies that the decay of the modes is slower, and for the RN-dSQ, $-Im$ $w$ increases with the increasing $q$. Therefore for the same parameter space, as the increasing magnetic charge $q$, RN-dSQ is more stable than the Bardeen-Kiselev BH with cosmological constant. In Fig.\ref{fig3} and Fig.\ref{fig4}, we observed the similar qualitative behaviours of Bardeen-Kiselev BH with cosmological constant and RN-dSQ, i.e. both the oscillation frequency and the damping rate decrease with increasing values of $c$ and $\lambda$.

From Fig.\ref{fig2} to Fig.\ref{fig4}, it shows that the oscillation frequency and the damping rate of Bardeen-Kiselev BH with cosmological constant and RN-dSQ increase with the increasing $l$ and $w$, and considering the effect of the state parameter $w$ and the normalization constant $c$ on the oscillation frequency and the damping rate of Bardeen-Kiselev BH with cosmological constant, we can conclude that, due to the presence of quintessence, the gravitational perturbations of the Bardeen-Kiselev BH with cosmological constant damp more slowly and oscillate more slowly and this result is consistent with the Ref.\cite{retMSA}. Moreover, by varying the BH parameters, we analyse the behaviour of both real and imaginary parts of Bardeen-Kiselev quasinormal frequencies and compare frequencies with RN-dSQ. Interestingly, it should be noted that for the gravitational perturbations, the response of Bardeen-Kiselev BH with cosmological constant and RN-dSQ in terms of the imaginary part of $w$ are different only when the charge parameter $q$ is varied, and this behavior can be used to understand nonlinear and linear electromagnetic fields in curved spacetime separately.

In table \ref{tab1}, table \ref{tab2} and table \ref{tab3}, we report the quasinormal frequencies in electromagnetic perturbations of Bardeen-Kiselev BH with cosmological constant and compare frequencies with RN-dSQ. Concretely speaking, in table \ref{tab1} we present the QNM frequencies of electromagnetic field perturbation for Bardeen-Kiselev BH with cosmological constant and RN-dSQ in terms of the changeable magnetic charge $q$, it shows that the response of the QNM frequencies of Bardeen-Kiselev BH with cosmological constant and RN-dSQ under electromagnetic perturbations are same with the gravitational perturbation when the charge parameter $q$ is changeable. We present the QNM frequencies of electromagnetic perturbations for Bardeen-Kiselev BH with cosmological constant and RN-dSQ in terms of the changeable normalization factor $c$ in table \ref{tab2}, it shows that the oscillation frequency of the two BHs decrease with the increasing values of $c$, while for the Bardeen-Kiselev BH with cosmological constant, the damping rate increases with the increasing values of $c$, which implies the decay of the modes for Bardeen-Kiselev BH with cosmological constant is faster, and the damping rate of RN-dSQ decrease with the increasing values of $c$.  In table \ref{tab3} we present the QNM frequencies of electromagnetic field perturbation for Bardeen-Kiselev BH with cosmological constant and RN-dSQ in terms of the changeable cosmological constant $\lambda$, which shows that both the oscillation frequency and the damping rate decrease with the increasing values of $\lambda$, and this behavior is same with the gravitational perturbations. Through these tables, for the same BH parameters, when the state parameter $w$ increases, the real parts of the quasinormal frequencies of Bardeen-Kiselev BH with cosmological constant increase while the absolute values of the imaginary parts decrease, while the real parts of the quasinormal frequencies and the absolute values of the imaginary parts of RN-dSQ increase, also considering the effect of the state parameter $w$ and the normalization constant $c$ on the oscillation frequency and the damping rate of Bardeen-Kiselev BH with cosmological constant, we can conclude that, in the presence of quintessence, the electromagnetic perturbations of the Bardeen-Kiselev BH with cosmological constant damp more faster and and oscillate more slowly, this behavior is different with the gravitational perturbation.

Regarding these figures and tables, by varying the BH parameters, we analyse the behaviour of both real and imaginary parts of Bardeen-Kiselev quasinormal frequencies and compare frequencies with RN-dSQ. Interestingly, it shows that the response of Bardeen-Kiselev BH with cosmological constant and RN-dSQ under electromagnetic perturbations are different when the charge parameter $q$, the state parameter $w$ and the normalization factor $c$ are varied, but for the gravitational perturbations, the response of Bardeen-Kiselev BH with cosmological constant and RN-dSQ are different only when the charge parameter $q$ is varied. Therefore, we may view that compared with the gravitational perturbation, the electromagnetic perturbation can be used to understand nonlinear and linear electromagnetic fields in curved spacetime separately.

\begin{figure}[htbp]
\begin{tabular}{cc}
\begin{minipage}[t]{0.45\linewidth}
\centerline{\includegraphics[width=6.0cm]{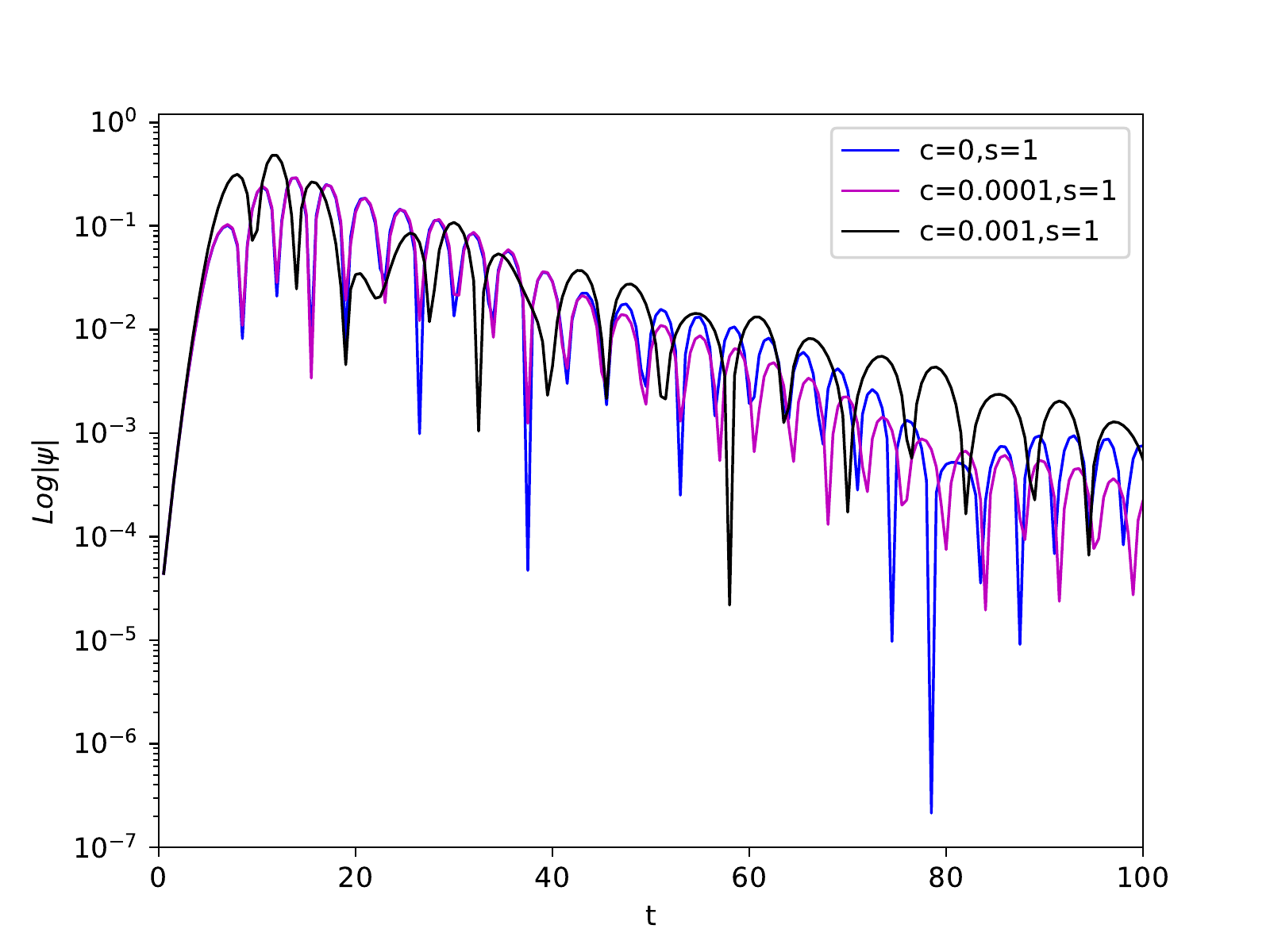}}
\centerline{(a)}
\end{minipage}
\hspace{7mm}
\begin{minipage}[t]{0.45\linewidth}
\centerline{\includegraphics[width=6.0cm]{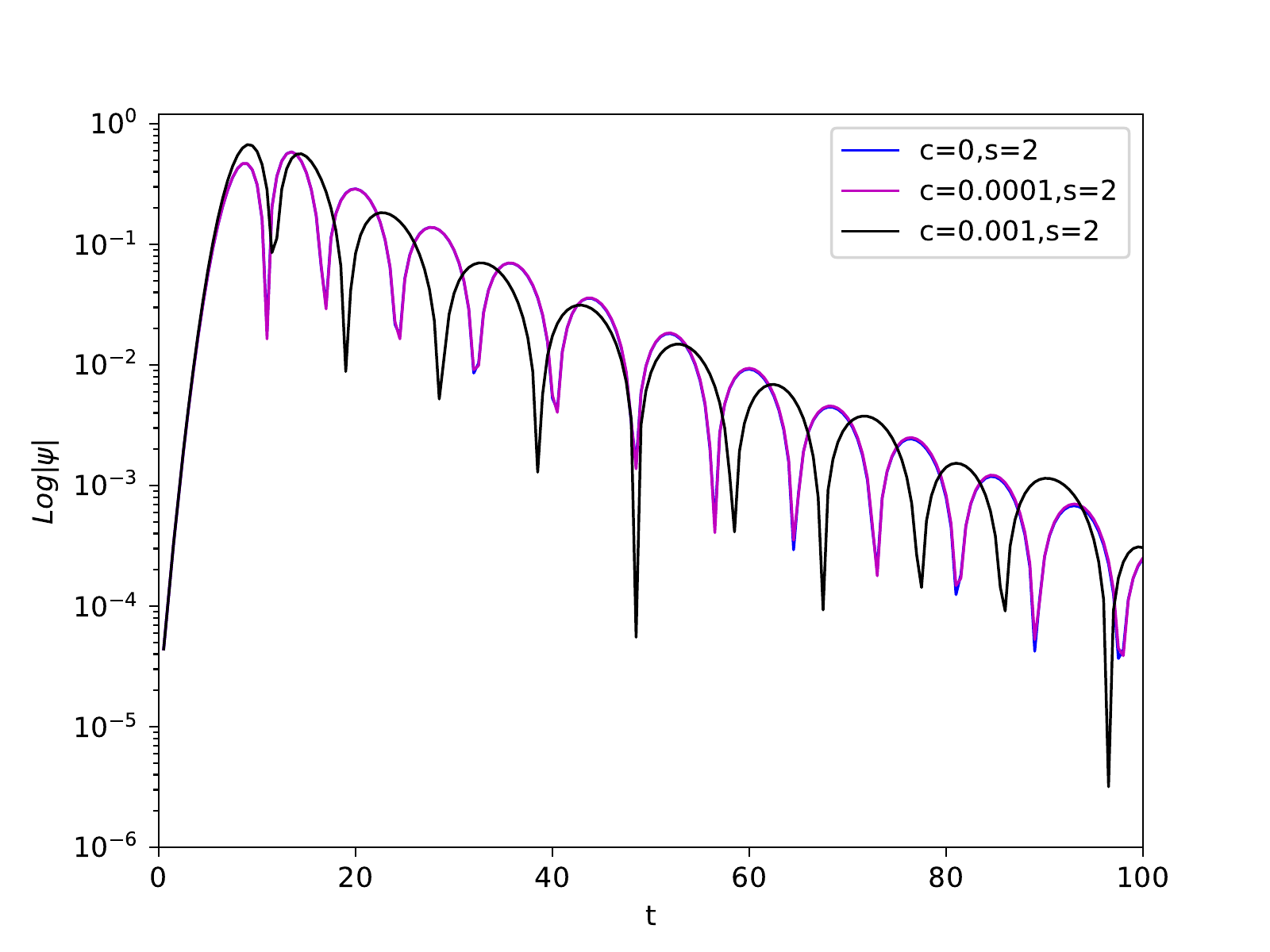}}
\centerline{(b)}
\end{minipage}
\end{tabular}
\caption{(a)The dynamical evolution of nonlinear electrodynamics field in the background of the Bardeen-Kiselev BH spacetime$(S=1,\ell=3)$; (b) The dynamical evolution of gravitational perturbation in the background of the Bardeen-Kiselev BH spacetime$(S=2,\ell=2)$. In both cases we take $M=1$, $q=0.3$, $w=-2/3$ and $\lambda=0.0003$.}
\label{fig14}
\end{figure}

Finally, regarding to the QNMs oscillation shape, we use the finite difference method to study the dynamical evolution of the nonlinear electrodynamics field perturbations and gravitational perturbations in the time domain \cite{retJLI,retSRW} and examine the stability of Bardeen-Kiselev BH with cosmological constant. As seen in Fig.\ref{fig14}, it shows the evolution of perturbation in log scale. We can see that the QNM oscillation of electromagnetic perturbations for Bardeen-Kiselev BH with cosmological constant is sensitive to the smaller normalization factor $c$, also when the normalization factor $c$ takes smaller values, the QNM oscillation of gravitational perturbations for Bardeen-Kiselev BH with cosmological constant and the absence of quintessence are almost the same, therefore which is not sensitive to the smaller values $c$. While for the gravitational perturbations, as the normalization factor $c$ increases, the oscillation frequency will decrease, this result is consistent with the result obtained by WKB method. Moreover, in view of the finite value of cosmological horizon $r_{c}$, our integration domain is limited in the range of $-r_{*}$ to $r_{*}$, it is tricky to obtain numerical values of $\varphi$ at late time, thus there is no any power law tail in the dynamics, and if we observe the Eq.\eqref{eq5}, it is found that as $\lambda$ $\rightarrow$ 0,  $r_{c}$ $\rightarrow$ $\infty$, therefore numerically domain of integration also becomes large enough and finally one can obtain very late time dynamics, but in the fact that this small value of $\lambda$ in numerical computation has its own challenges.

{\centering  \section{Greybody factors and absorption coefficients} \label{secV} }

By using the WKB method to obtain reflection and transmission coefficients(greybody factors) has been found in the literatures\cite{ret33,ret34,ret35}, such as in the context of  braneworld models and wormholes so on. In this section we intend to discuss the frequency dependent reflection $R(\omega)$ and transmission coefficient $T(\omega)$ for a scattering process of the gravitational and electromagnetic perturbations from the Bardeen-Kiselev BH with cosmological constant. According to the Hawking radiation\cite{ret36}, at the event horizon, the emission rate of BH in a mode with frequency $\omega$ is
\begin{equation}
\Gamma(\omega)=\frac{1}{\mathrm{e}^{\alpha \omega} \pm 1} \frac{\mathrm{d}^3 K}{(2 \pi)^3}, \label{eq28}
\end{equation}
where $\alpha$ represents the inverse of Hawking temperature and the symbol $\pm$ corresponds to fermions (bosons). However, not all the radiation are able to reach the distant observer, i.e. a part of the radiation would be tunneled through the potential barrier and reach the distant observer while the other part would be reflected back towards the BH, and the radiation recorded by the distant observer will no longer appear as a blackbody. Therefore, the emission rate measured by an observer at infinity for a frequency mode $\omega$ can be expressed as
\begin{equation}
\Gamma(\omega)=\frac{\gamma_{l}}{\mathrm{e}^{\alpha \omega} \pm 1} \frac{\mathrm{d}^3 K}{(2 \pi)^3}, \label{eq29}   \end{equation}
here $\gamma_{l}$ is the gray-body factor, which is defined as
\begin{equation}
\gamma_{l}= \left| T{(\omega)} \right|^{2}. \label{eq30}
\end{equation}
The formalism of the asymptotic behaviour of the wave after scattering off of the effective potential can be expressed in tortoise coordinate as
\begin{equation}
\Psi(r_{*})=T(\omega)e^{-i\omega r_{*}}, r_{*} \rightarrow -\infty (r \rightarrow r_{+})                \label{eq31}    \end{equation}
\begin{equation}
\Psi(r_{*})=e^{-i\omega r_{*}}+R(\omega)e^{i\omega r_{*}}.  r_{*} \rightarrow +\infty (r \rightarrow r_{c}),   \label{eq32}    \end{equation}
as seen in Eq.\eqref{eq31} and Eq.\eqref{eq32}, the reflection and transmission coefficients are functions of oscillation frequency $\omega$ of the wave. The reflection coefficient in the presence of the WKB approximation is given by
\begin{equation}
R(\omega)=(1+e^{-2\pi i \eta})^{-\frac{1}{2}},               \label{eq33}    \end{equation}
where $\eta$ is
\begin{equation}
\eta= \frac{i(\omega^{2}-V(r_{0}))}{\sqrt{-2V^{''}(r_{0})}}-\varLambda_{i}, i=2,3            \label{eq34}    \end{equation}
noted that the values of $\varLambda_{i}$ can be found out from Eq.\eqref{eq27}, thus conserving probability we have
\begin{equation}
\gamma_{l}= 1-\left| R{(\omega)} \right|^{2}. \label{eq35}
\end{equation}

Next, by using the WKB method, we will describe the calculation of $R(\omega)$ and $T(\omega)$. Besides, if $r_{0}$ is the value of $r$ where the the potential $V({r})$ is the maximum, as seen in the Ref.\cite{ret37}, there are three cases to explain the relation between $\omega^{2}$ and $V({r_{0}})$, and we we will focus on the case $\omega^{2}$ $\sim$ $V({r_{0}})$, because the WKB approximation has high accuracy.

\begin{figure}[htbp]
\begin{tabular}{cc}
\begin{minipage}[t]{0.45\linewidth}
\centerline{\includegraphics[width=7.0cm]{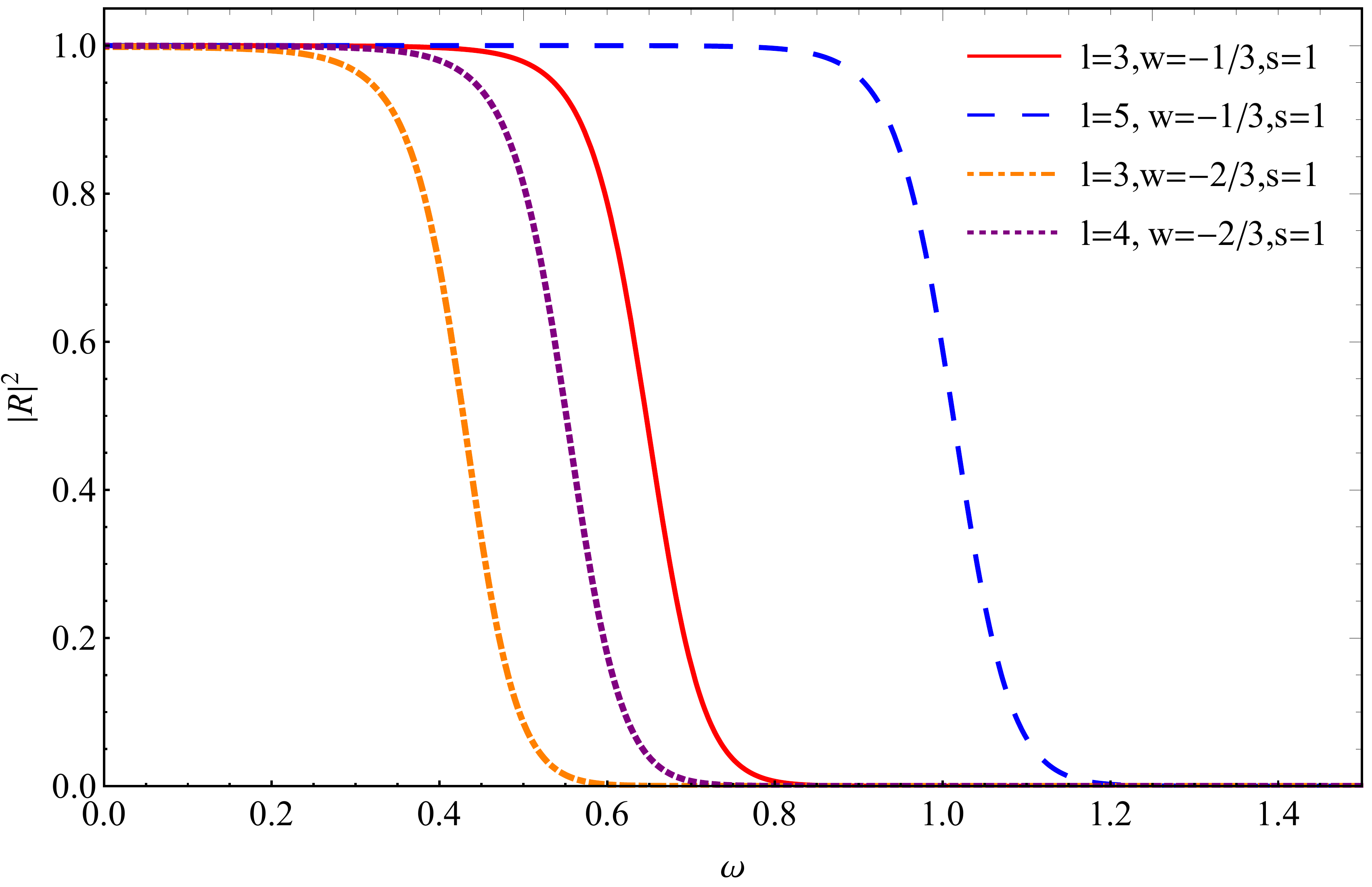}}
\centerline{(a)}
\end{minipage}
\hspace{7mm}
\begin{minipage}[t]{0.45\linewidth}
\centerline{\includegraphics[width=7.0cm]{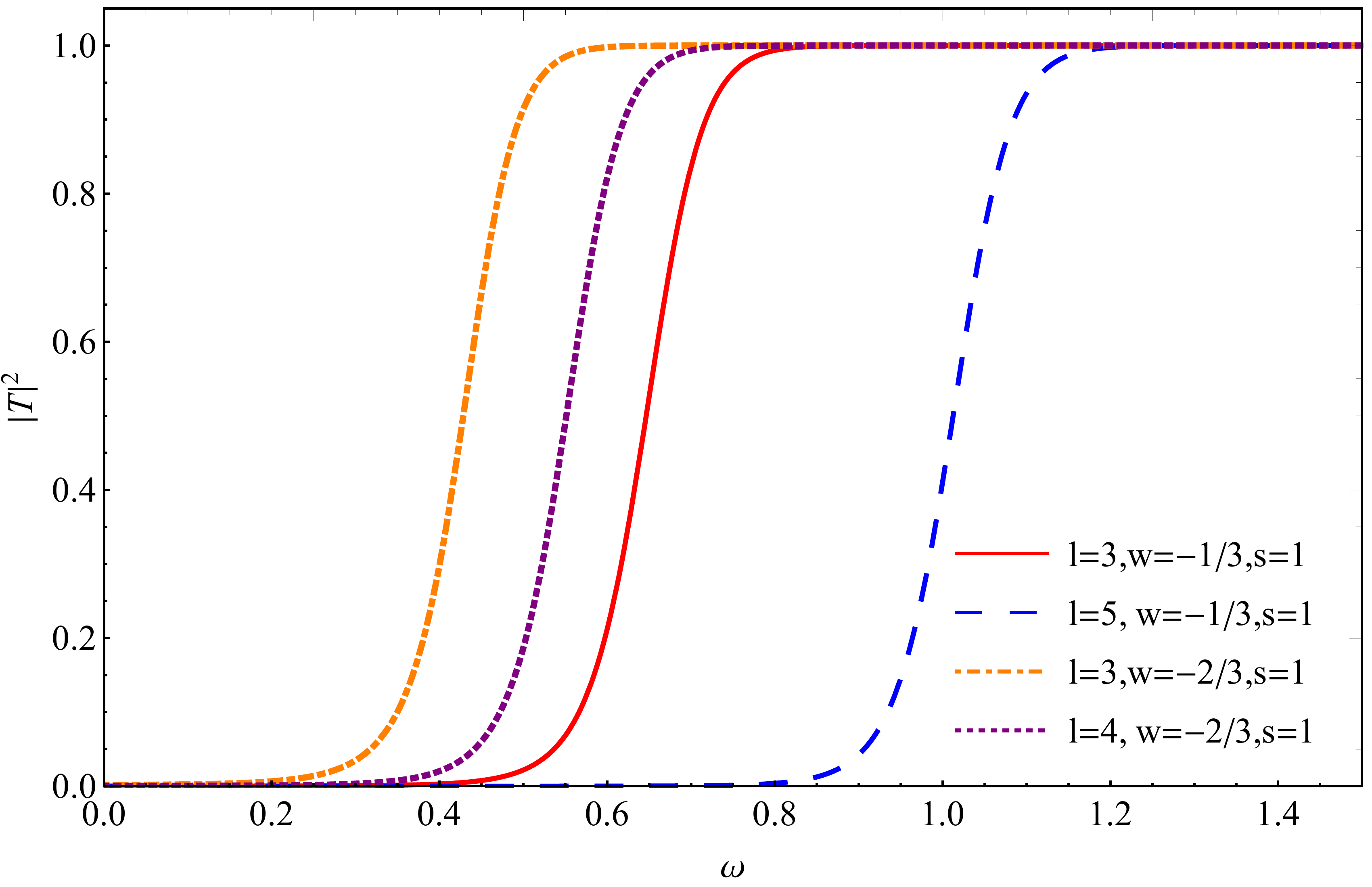}}
\centerline{(b)}
\end{minipage}
\\
\begin{minipage}[t]{0.45\linewidth}
\centerline{\includegraphics[width=7.0cm]{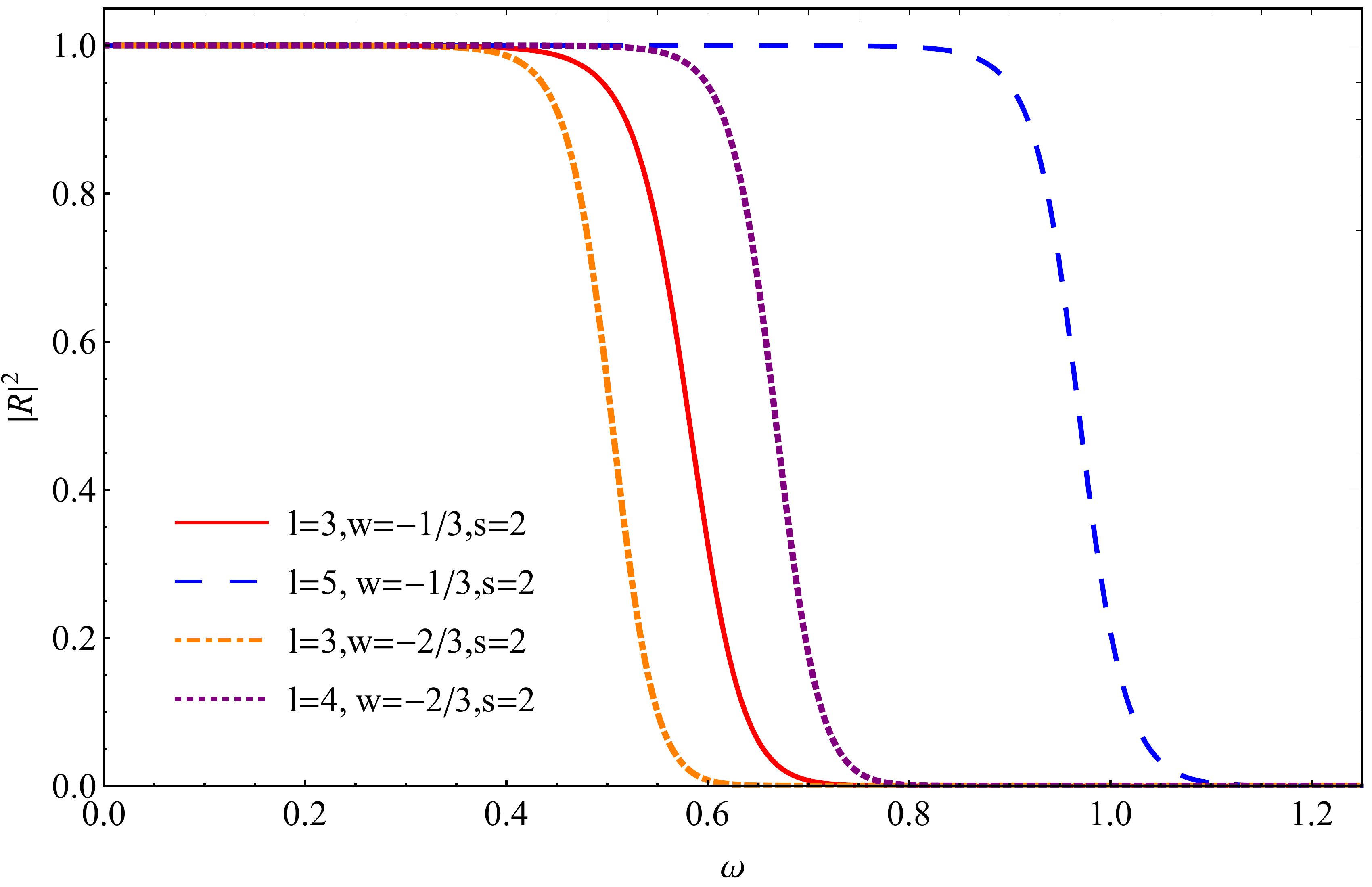}}
\centerline{(c)}
\end{minipage}
\hspace{7mm}
\begin{minipage}[t]{0.45\linewidth}
\centerline{\includegraphics[width=7.0cm]{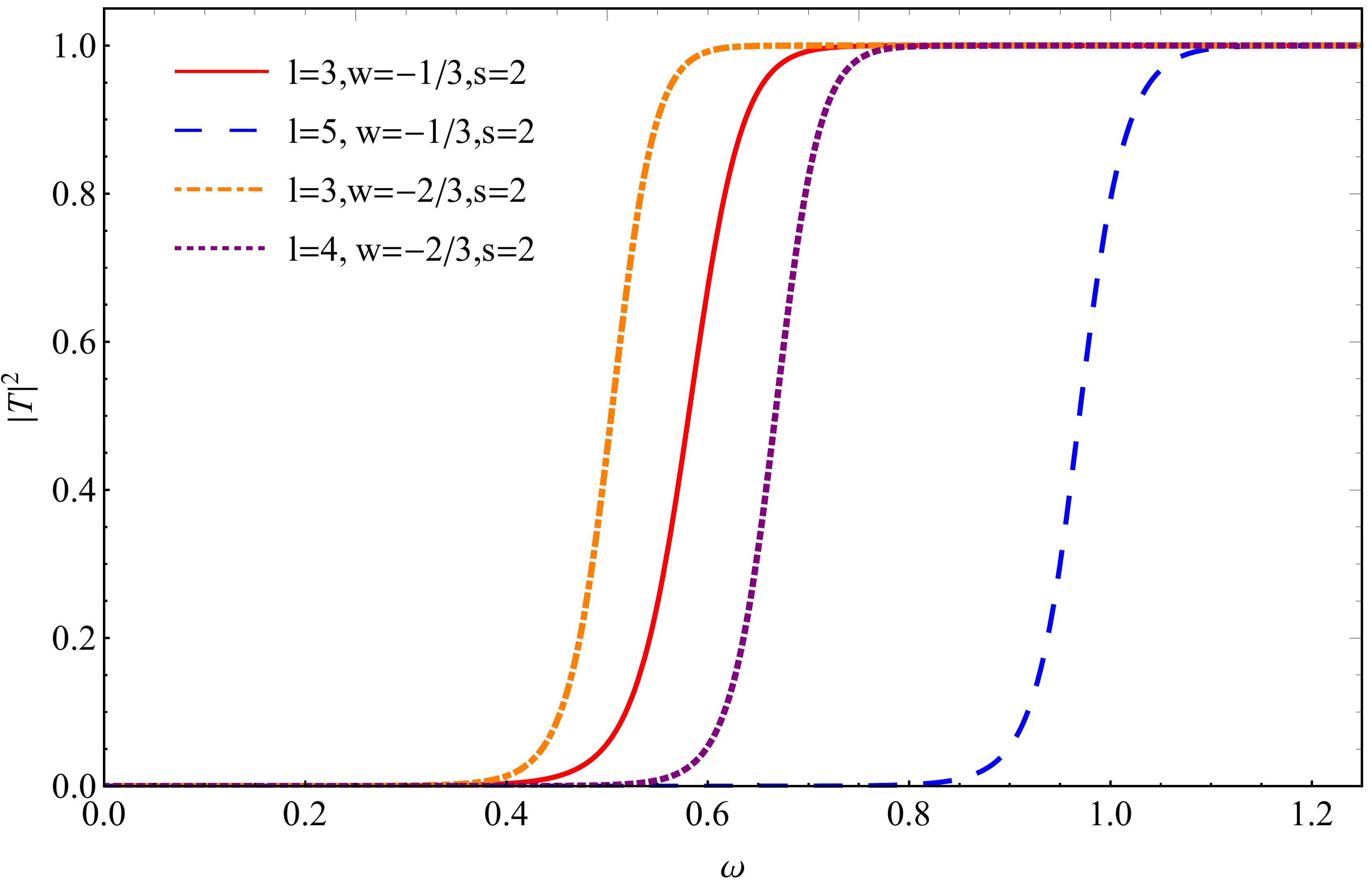}}
\centerline{(d)}
\end{minipage}
\end{tabular}
\caption{(a) $\lvert R\lvert ^{2}$ vs $\omega$ for electromagnetic perturbations; (b) $\lvert T\lvert ^{2}$ vs $\omega$ for electromagnetic perturbations; (c) $\lvert R\lvert ^{2}$ vs $\omega$ for gravitational perturbations; (d) $\lvert T\lvert ^{2}$ vs $\omega$ for gravitational perturbations. In both cases we take $M=1$, $q=0.25$,  $\lambda=0.0025$, $c=0.02$, and electromagnetic perturbations ($s=1$), gravitational perturbations($s=2$).}
\label{fig5}
\end{figure}

\begin{figure}[htbp]
\begin{tabular}{cc}
\begin{minipage}[t]{0.45\linewidth}
\centerline{\includegraphics[width=7.0cm]{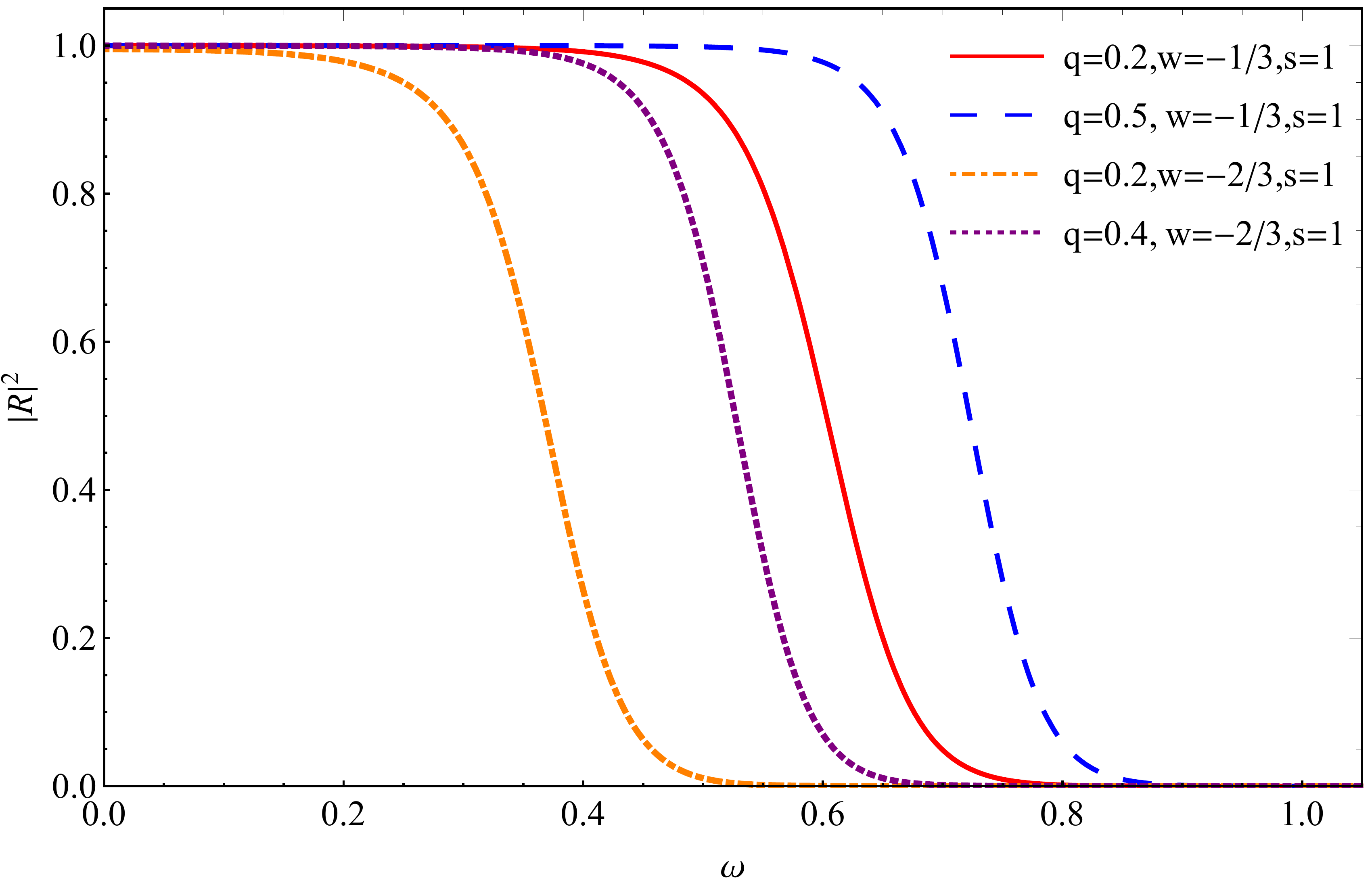}}
\centerline{(a)}
\end{minipage}
\hspace{7mm}
\begin{minipage}[t]{0.45\linewidth}
\centerline{\includegraphics[width=7.0cm]{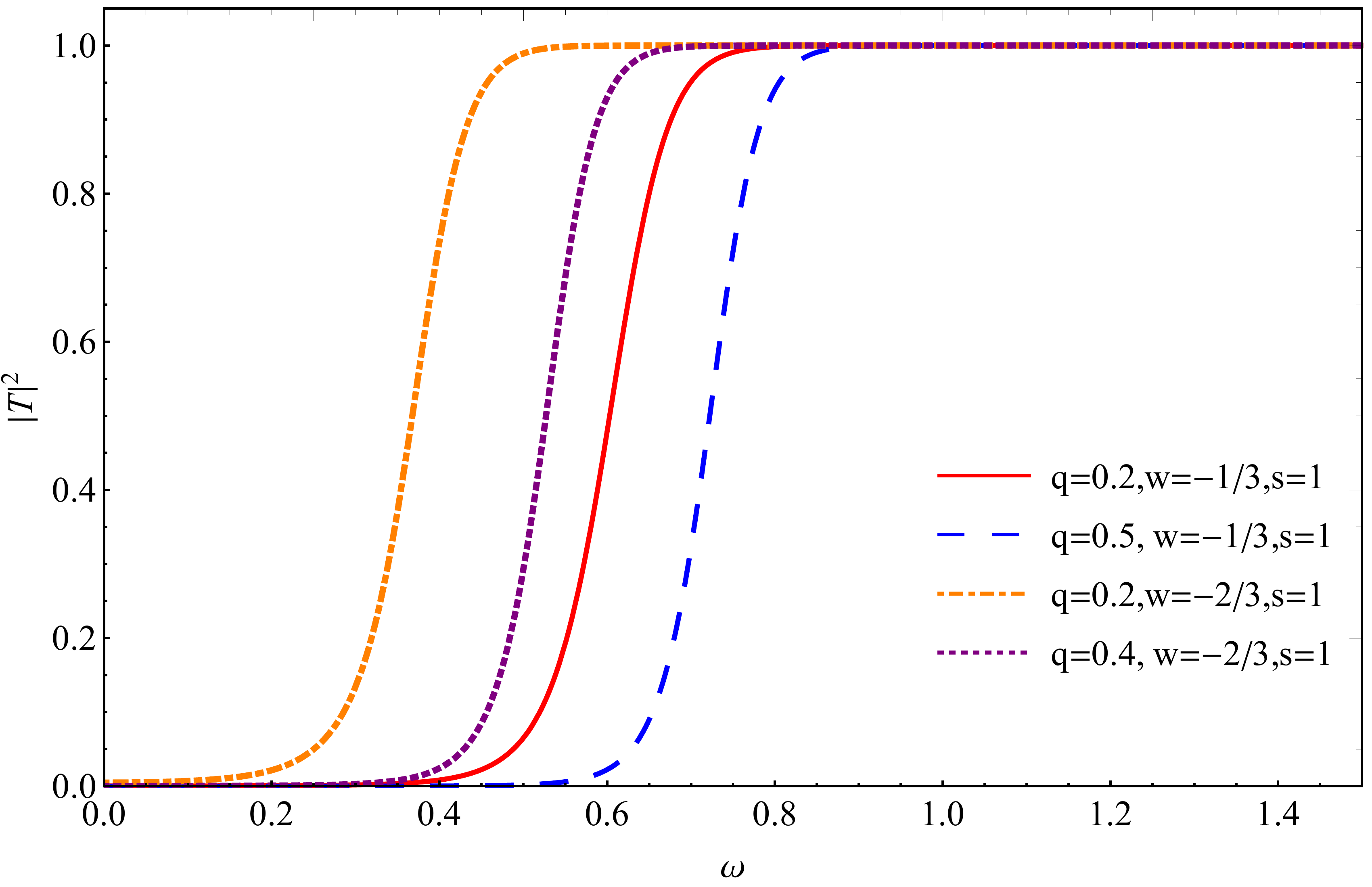}}
\centerline{(b)}
\end{minipage}
\\
\begin{minipage}[t]{0.45\linewidth}
\centerline{\includegraphics[width=7.0cm]{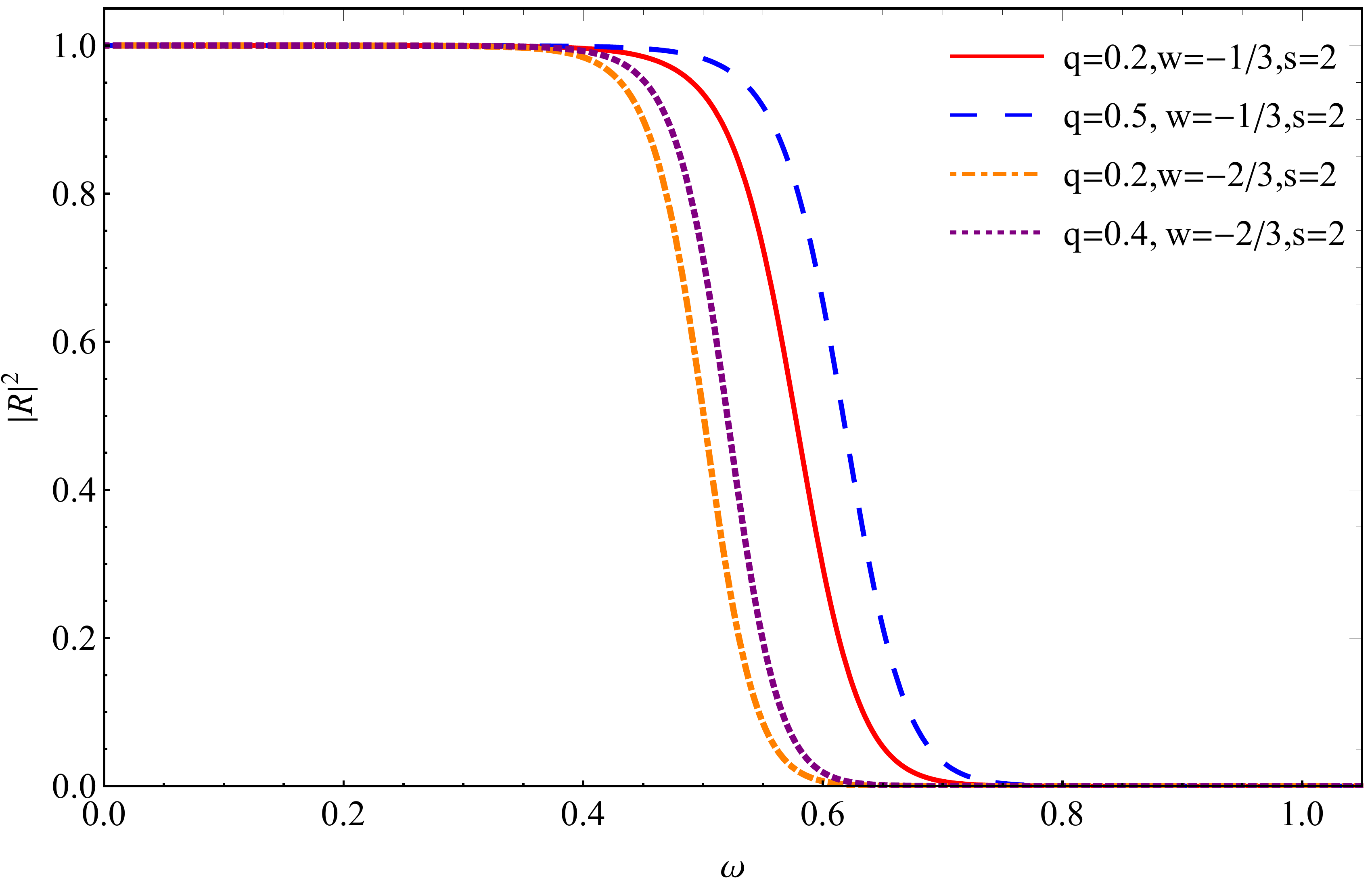}}
\centerline{(c)}
\end{minipage}
\hspace{7mm}
\begin{minipage}[t]{0.45\linewidth}
\centerline{\includegraphics[width=7.0cm]{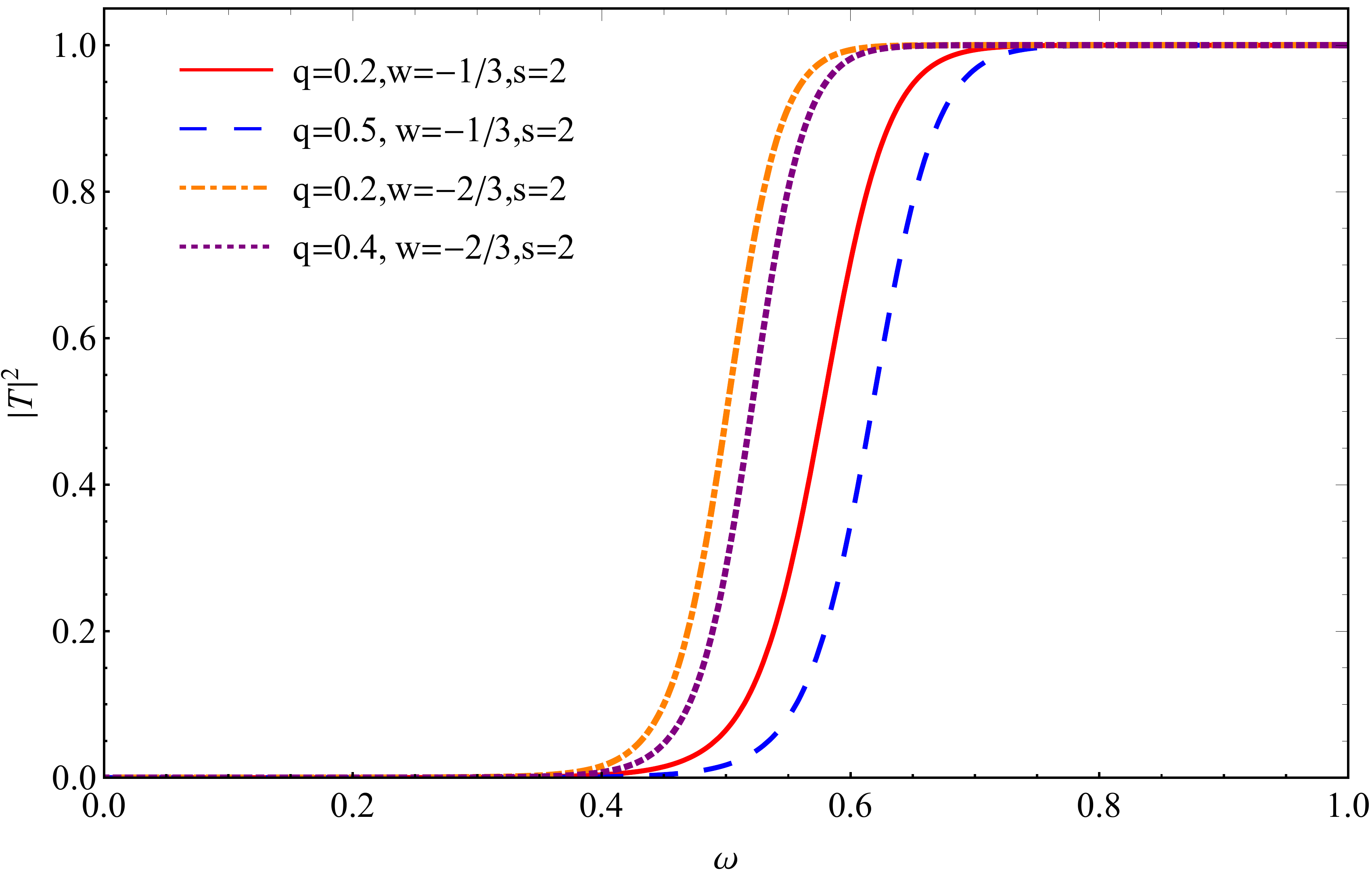}}
\centerline{(d)}
\end{minipage}
\end{tabular}
\caption{(a) $\lvert R\lvert^{2}$ vs $\omega$ for electromagnetic perturbations; (b) $\lvert T\lvert^{2}$ vs $\omega$ for electromagnetic perturbations; (c) $\lvert R\lvert^{2}$ vs $\omega$ for gravitational perturbations; (d) $\lvert T\lvert^{2}$ vs $\omega$ for gravitational perturbations. In both cases we take $M=1$, $\ell=3$,  $\lambda=0.0025$, $c=0.02$, and electromagnetic perturbations ($s=1$), gravitational perturbations($s=2$).}
\label{fig6}
\end{figure}

\begin{figure}[htbp]
\begin{tabular}{cc}
\begin{minipage}[t]{0.45\linewidth}
\centerline{\includegraphics[width=7.0cm]{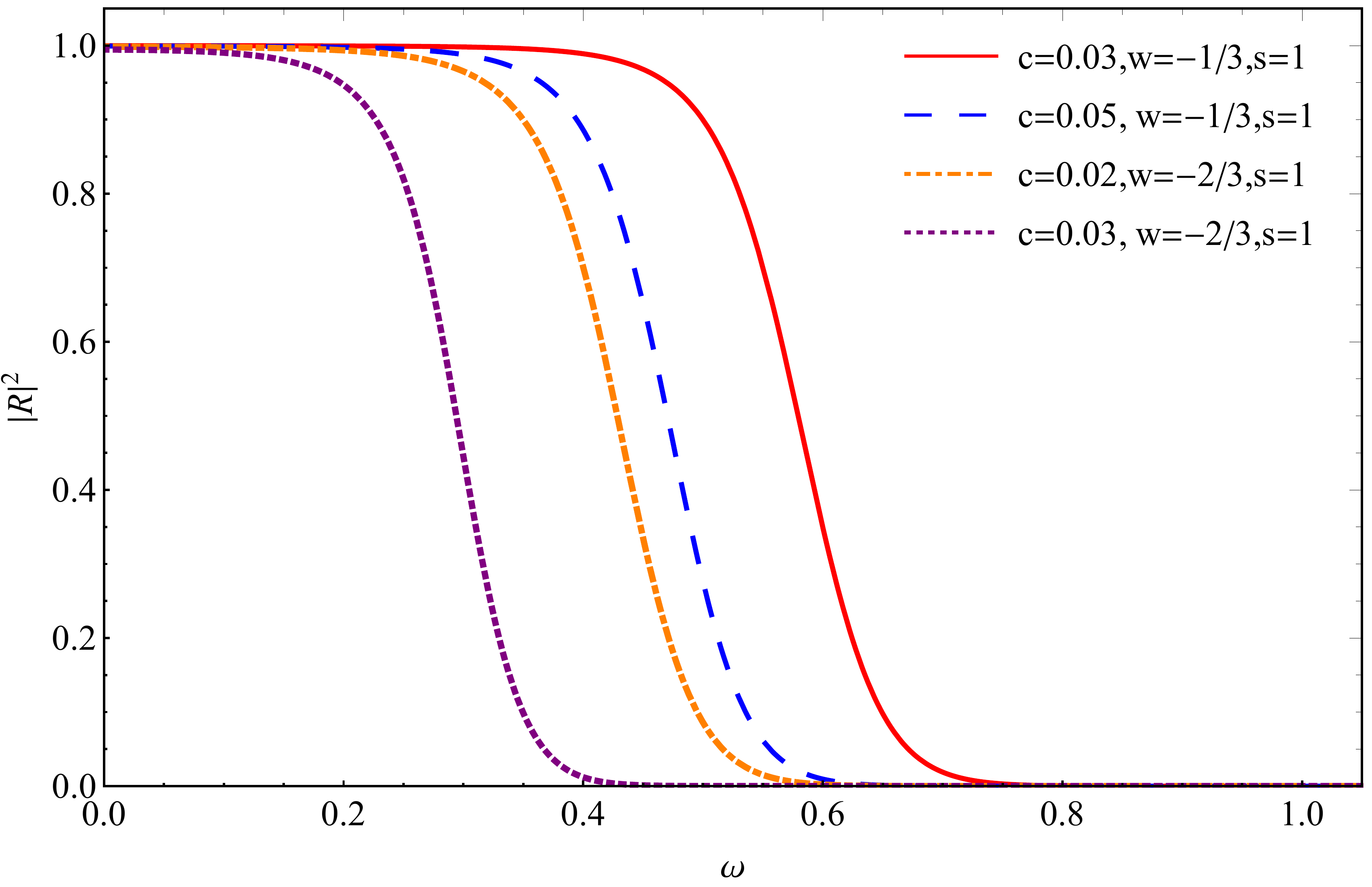}}
\centerline{(a)}
\end{minipage}
\hspace{7mm}
\begin{minipage}[t]{0.45\linewidth}
\centerline{\includegraphics[width=7.0cm]{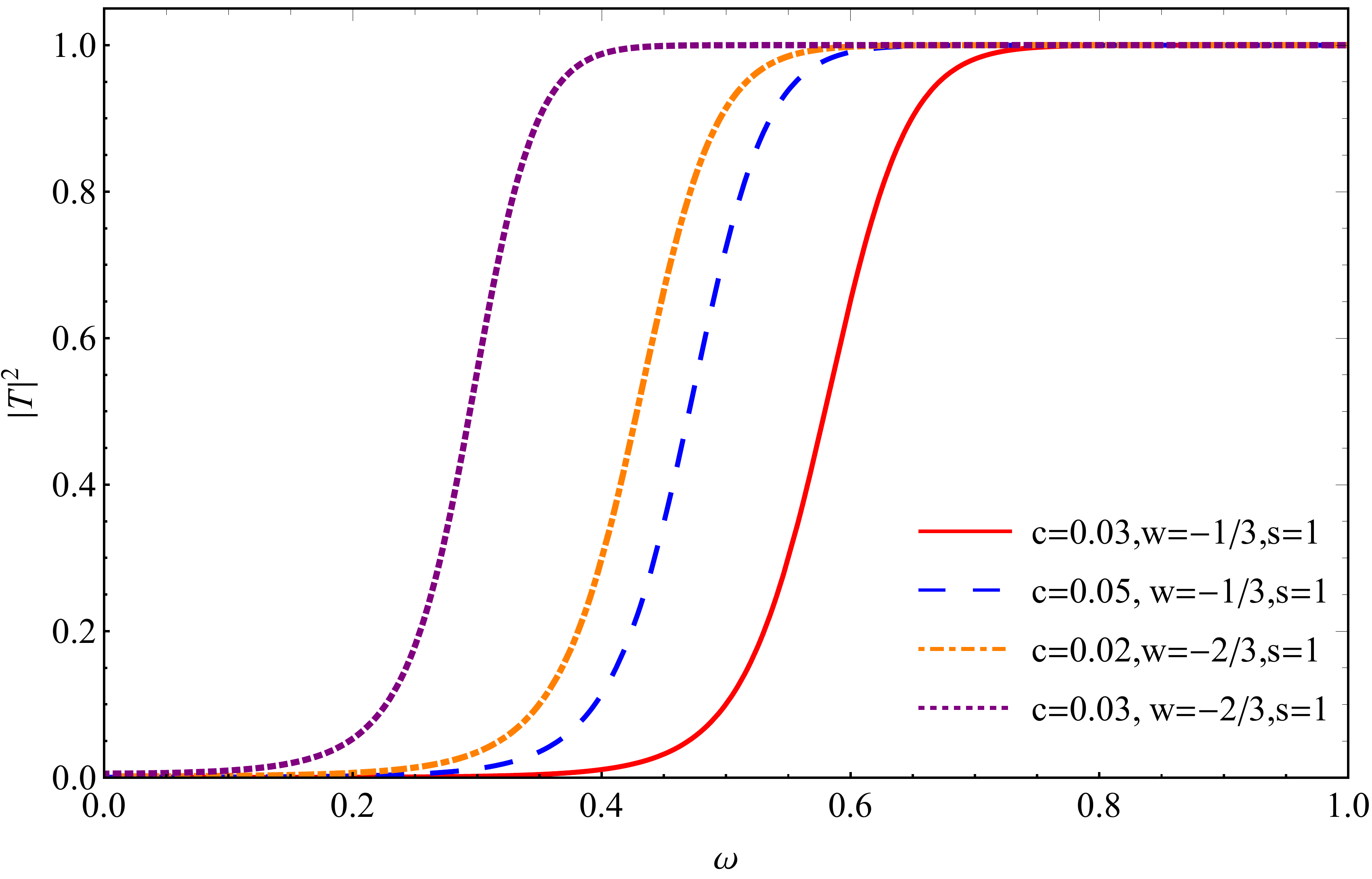}}
\centerline{(b)}
\end{minipage}
\\
\begin{minipage}[t]{0.45\linewidth}
\centerline{\includegraphics[width=7.0cm]{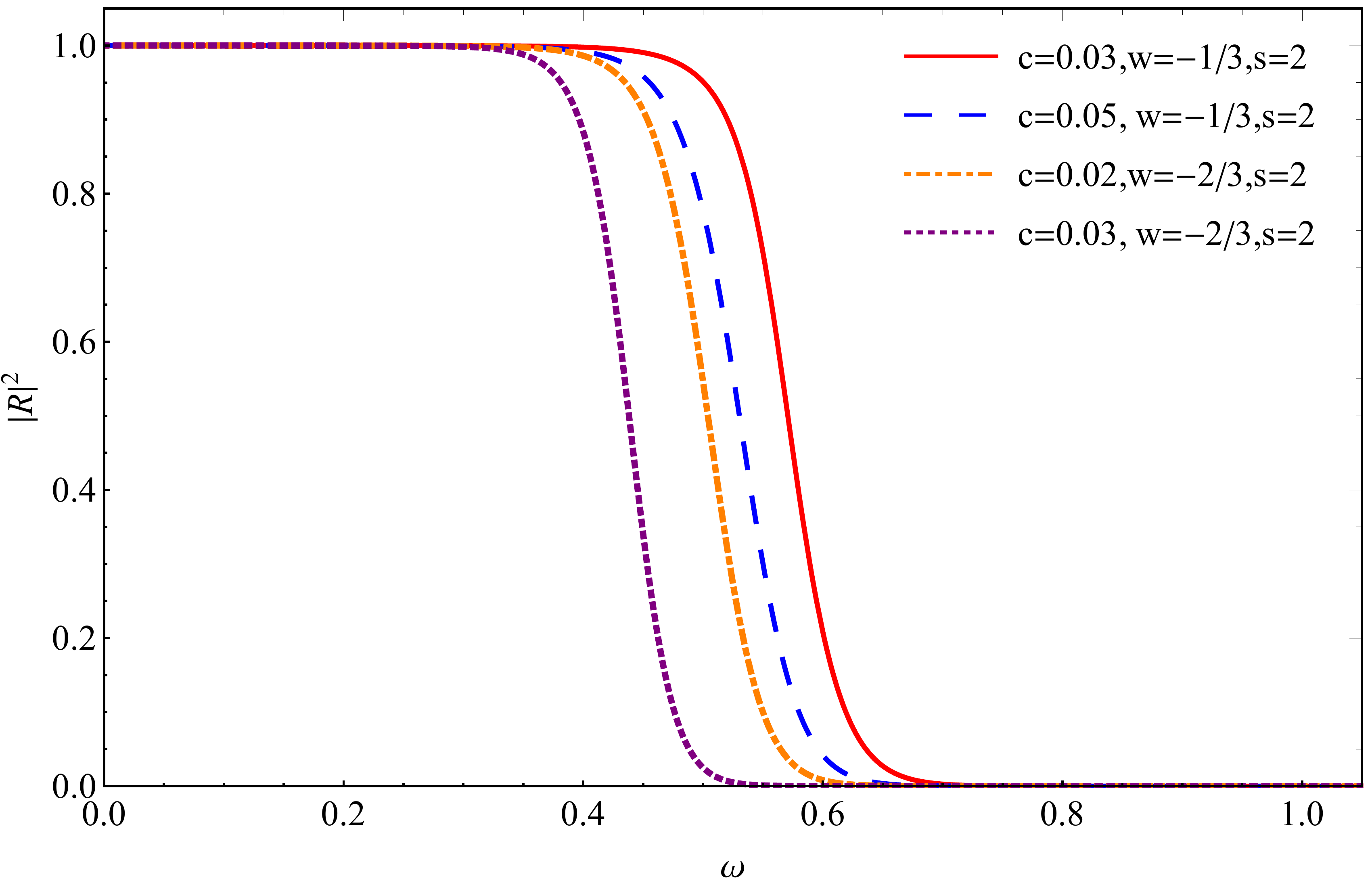}}
\centerline{(c)}
\end{minipage}
\hspace{7mm}
\begin{minipage}[t]{0.45\linewidth}
\centerline{\includegraphics[width=7.0cm]{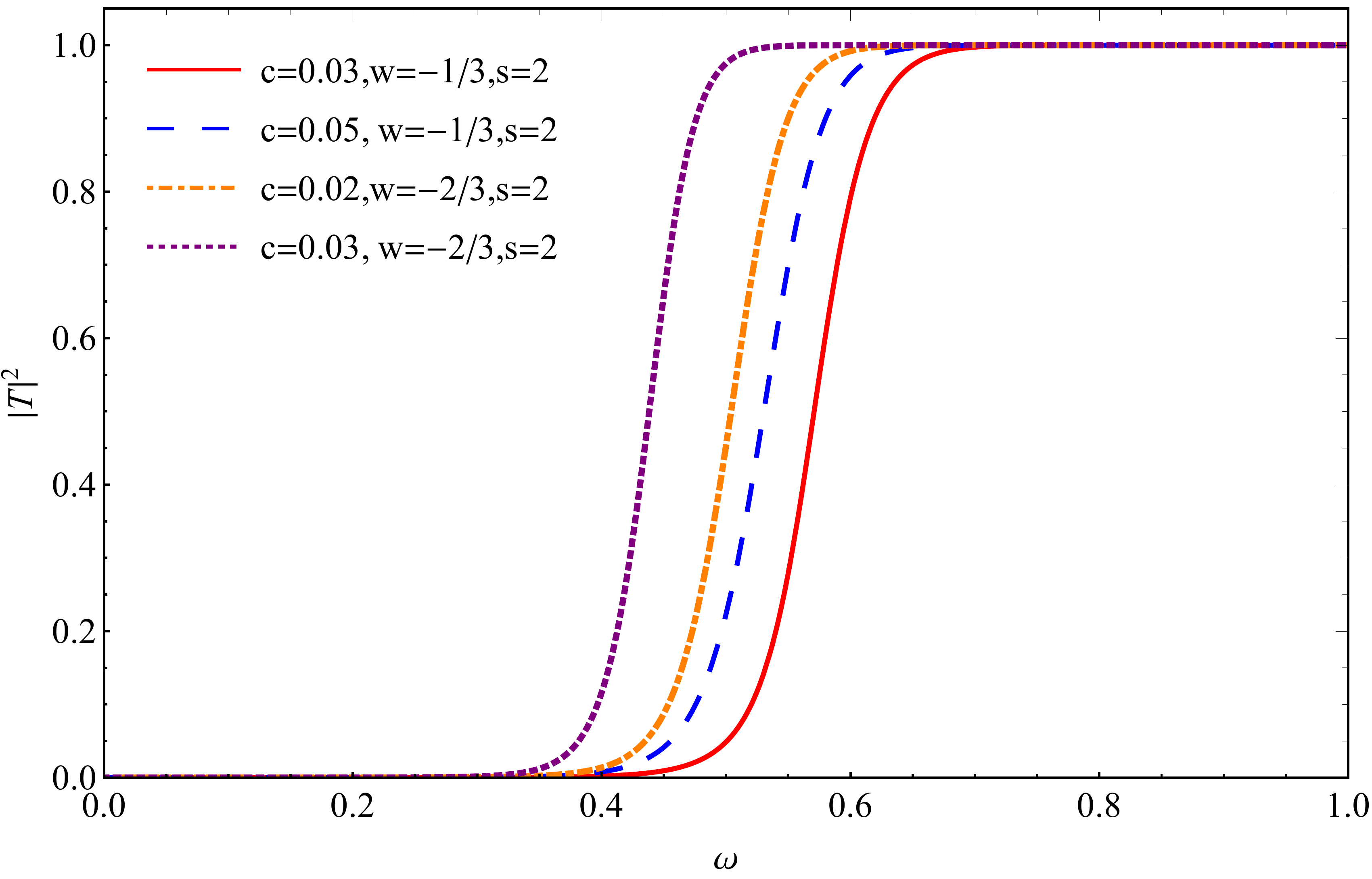}}
\centerline{(d)}
\end{minipage}
\end{tabular}
\caption{(a) $\lvert R\lvert^{2}$ vs $\omega$ for electromagnetic perturbations; (b) $\lvert T\lvert^{2}$ vs $\omega$ for electromagnetic perturbations; (c) $\lvert R\lvert^{2}$ vs $\omega$ for gravitational perturbations; (d) $\lvert T\lvert^{2}$ vs $\omega$ for gravitational perturbations. In both cases we take $M=1$, $\ell=3$,  $\lambda=0.0025$, $q=0.25$, and electromagnetic perturbations ($s=1$), gravitational perturbations($s=2$).}
\label{fig7}
\end{figure}

\begin{figure}[htbp]
\begin{tabular}{cc}
\begin{minipage}[t]{0.45\linewidth}
\centerline{\includegraphics[width=7.0cm]{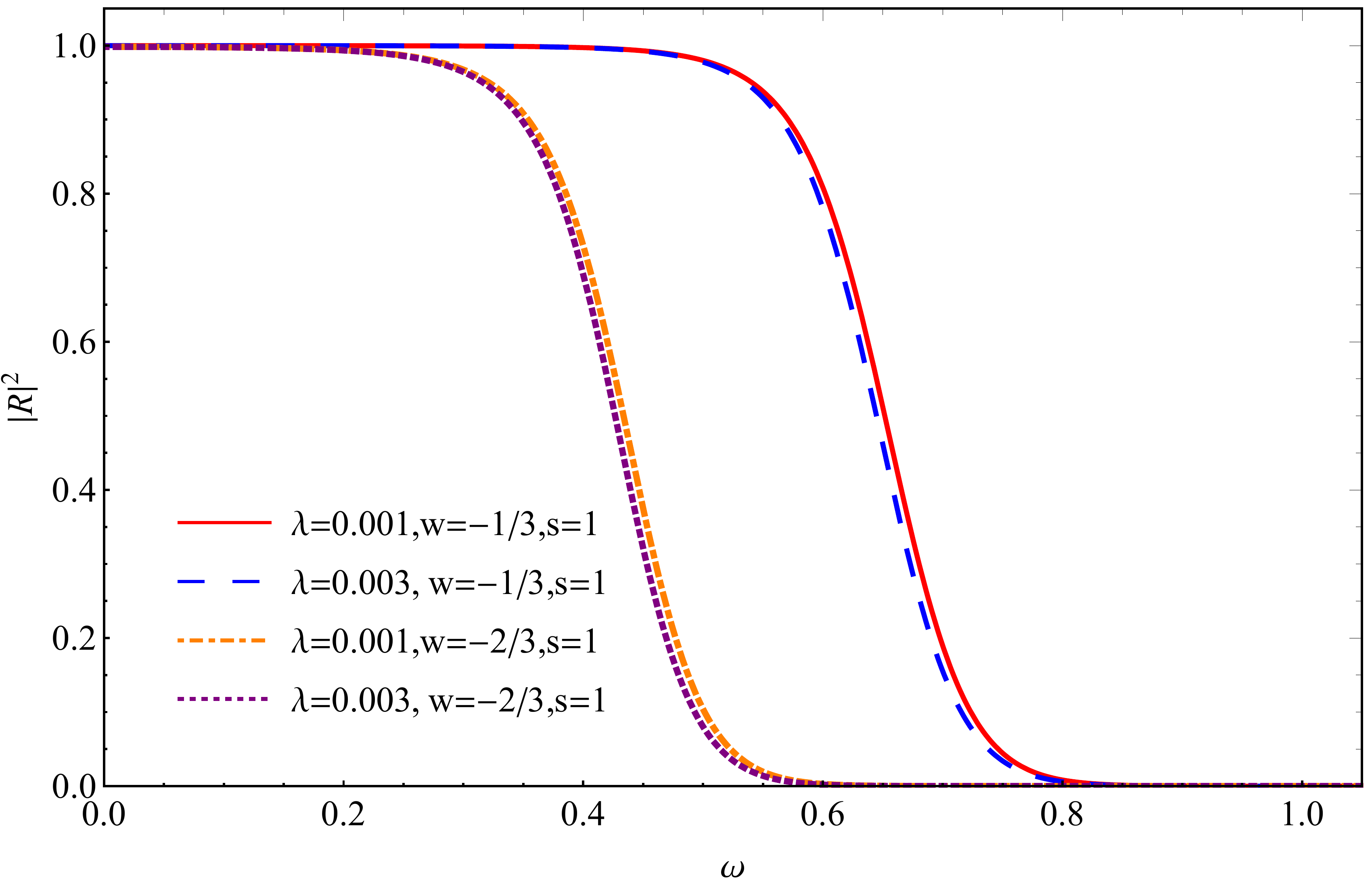}}
\centerline{(a)}
\end{minipage}
\hspace{7mm}
\begin{minipage}[t]{0.45\linewidth}
\centerline{\includegraphics[width=7.0cm]{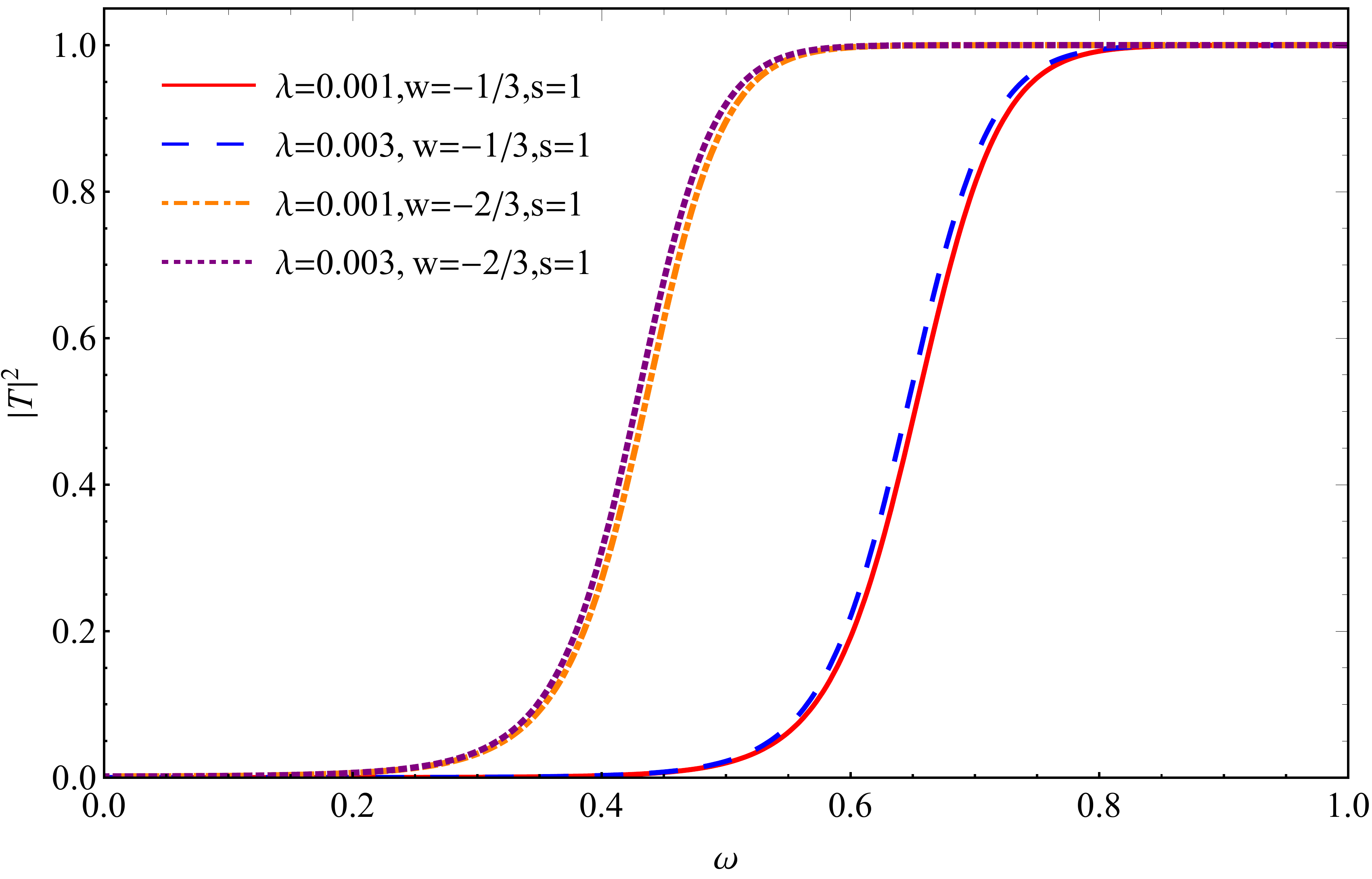}}
\centerline{(b)}
\end{minipage}
\\
\begin{minipage}[t]{0.45\linewidth}
\centerline{\includegraphics[width=7.0cm]{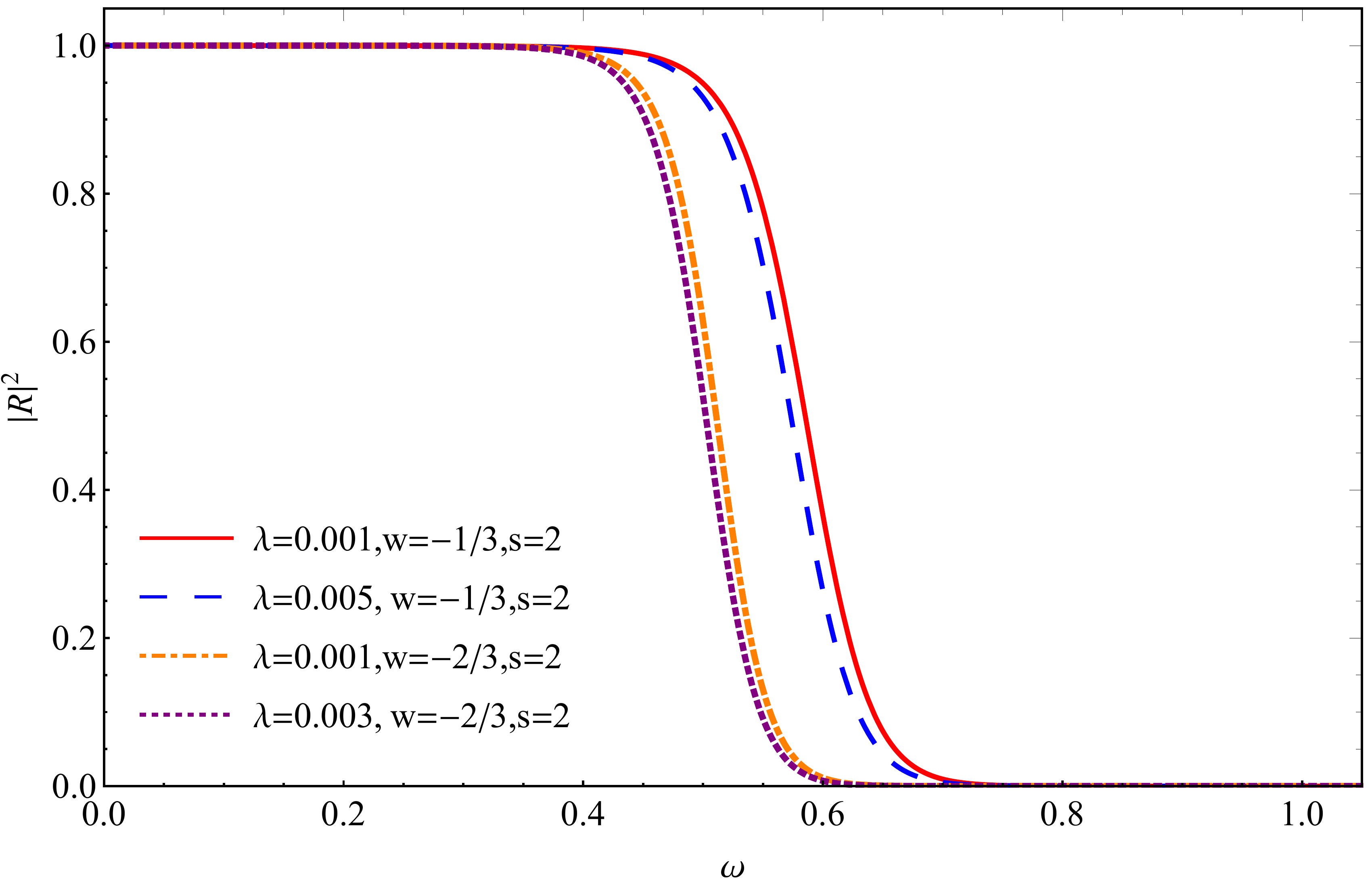}}
\centerline{(c)}
\end{minipage}
\hspace{7mm}
\begin{minipage}[t]{0.45\linewidth}
\centerline{\includegraphics[width=7.0cm]{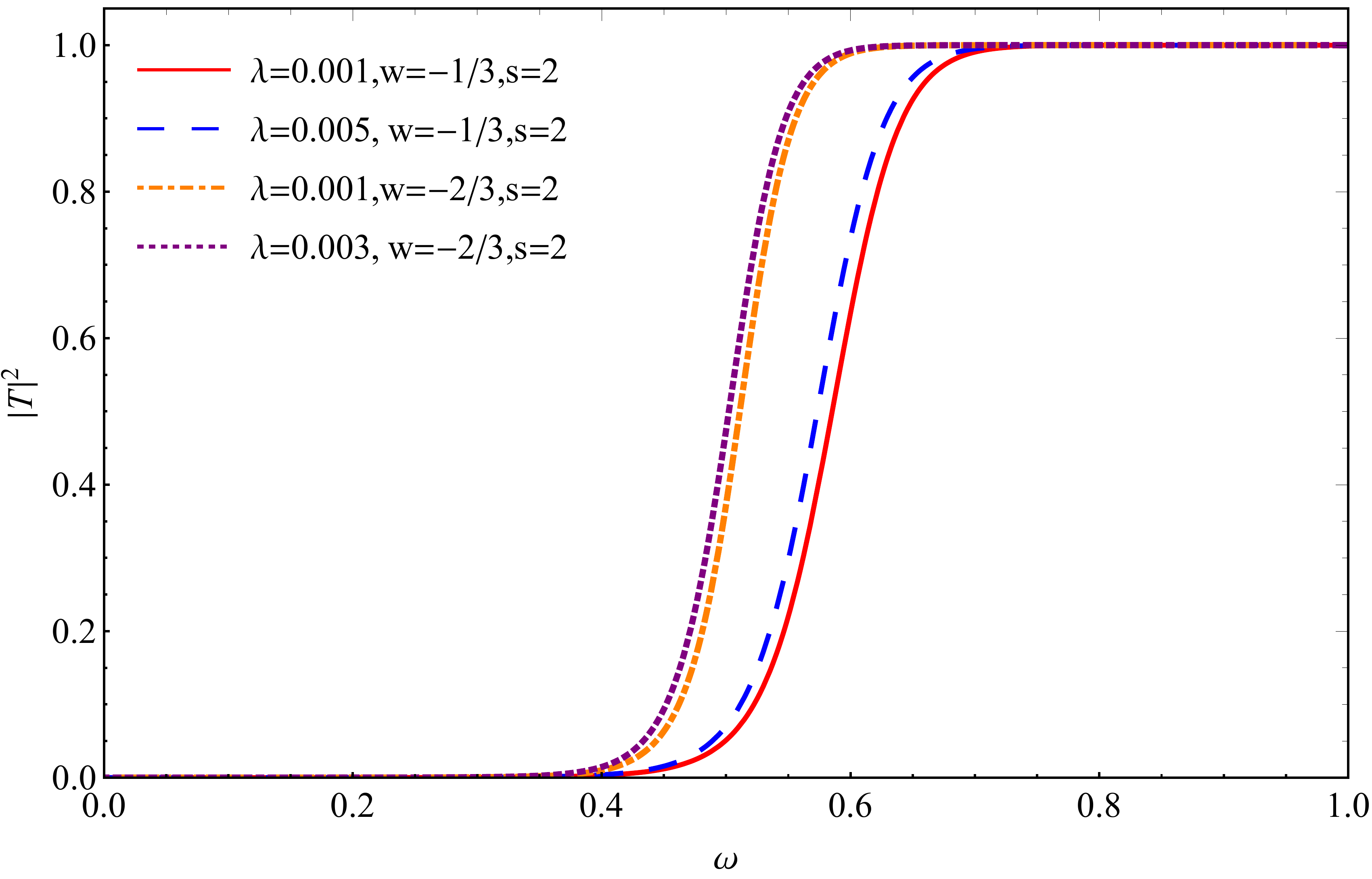}}
\centerline{(d)}
\end{minipage}
\end{tabular}
\caption{(a) $\lvert R\lvert^{2}$ vs $\omega$ for electromagnetic perturbations; (b) $\lvert T\lvert^{2}$ vs $\omega$ for electromagnetic perturbations; (c) $\lvert R\lvert^{2}$ vs $\omega$ for gravitational perturbations; (d) $\lvert T\lvert^{2}$ vs $\omega$ for gravitational perturbations. In both cases we take $M=1$, $\ell=3$,  $q=0.25$, $c=0.02$, and electromagnetic perturbations ($s=1$), gravitational perturbations($s=2$).}
\label{fig8}
\end{figure}

By using the effective potentials from Sec.\ref{secIII}, we numerically plot the variation of reflection and transmission coefficient on the frequency of electromagnetic perturbation and gravitational perturbation with various parameters($\ell$, $q$, $c$, $\lambda$, $w$) from Fig.\ref{fig5} to Fig.\ref{fig8}. Through these figures, we can see that the value of greybody bound is zero when the frequency is minimal, and the value of greybody bound turns out to be 1 when the frequency is large enough, which implies that the wave is basically totally reflected when the frequency is small, while in view of the tunneling effect, part wave could pass through the potential barrier when the frequency increases, or the wave will not be reflected when the frequency reaches a certain level. Moreover, noted that according to the Eq.\eqref{eq30} and Eq.\eqref{eq35}, for the high frequency and low frequency limits, the properties of Reflection coefficients and Transmission coefficients should be opposite, in fact these results can be verified from Fig.\ref{fig5} to Fig.\ref{fig8}.

The electromagnetic perturbations ($s=1$) and gravitational perturbations ($s=2$) in Fig.\ref{fig5} have similar behaviors, i.e. for a fixed frequency, the greybody factor will decrease with the increasing multipole number $\ell$, the response of greybody factor for different $\ell$ under gravitational perturbation is larger than the case under electromagnetic perturbation.

The similar behaviors for electromagnetic perturbations and gravitational perturbations appear in Fig.\ref{fig6}, for a fixed frequency, the greybody factor will decrease with the increasing magnetic charge $q$, this result is different with the Ref.\cite{retSA}, and it may be the reason that the existence of quintessence changes the metric (Eq.\eqref{eq5}). Besides, for the same spacetime parameters, the response of greybody factor for different $q$ under electromagnetic perturbations is larger than the case under gravitational perturbations.

Fig.\ref{fig7} shows that under electromagnetic perturbations and gravitational perturbations, for a fixed frequency, the greybody factor will increase with the increasing normalization factor $c$, and the response of greybody factor for different $c$ under electromagnetic perturbations is larger than the case under gravitational perturbations.

Fig.\ref{fig8} shows that similar behaviors for electromagnetic perturbations and gravitational perturbations, i.e. for a fixed frequency, the greybody factor will slightly increase with the increasing cosmological constant $\lambda$.

Through these figures, for the same parameters in terms of electromagnetic perturbations and gravitational perturbations, the greybody factor will decrease with the increasing state parameter $w$, and considering the effect of normalization factor $c$ on greybody factor in Fig.\ref{fig7}, we may view that due to the presence of quintessence, the transmission coefficient will increase. As mentioned above, the properties of reflection coefficients and transmission coefficients are opposite, then we will not present specific description for Reflection coefficients.

Finally, we will study the total absorption cross section in the context of electromagnetic and gravitational perturbations for different parameter spaces in the Bardeen-Kiselev BH with cosmological constant background. The total absorption cross section is given by
\begin{equation}
\begin{aligned}
\sigma_{\ell} &=\frac{\pi(2 \ell+1)}{\omega^{2}}\left|T_{\ell}(\omega)\right|^{2}, \\
\sigma &=\sum_{\ell} \frac{\pi(2 \ell+1)}{\omega^{2}}\left|T_{\ell}(\omega)\right|^{2} .   \label{eq33}
\end{aligned}
\end{equation}

\begin{figure}[htbp]
\begin{tabular}{cc}
\begin{minipage}[t]{0.45\linewidth}
\centerline{\includegraphics[width=7.0cm]{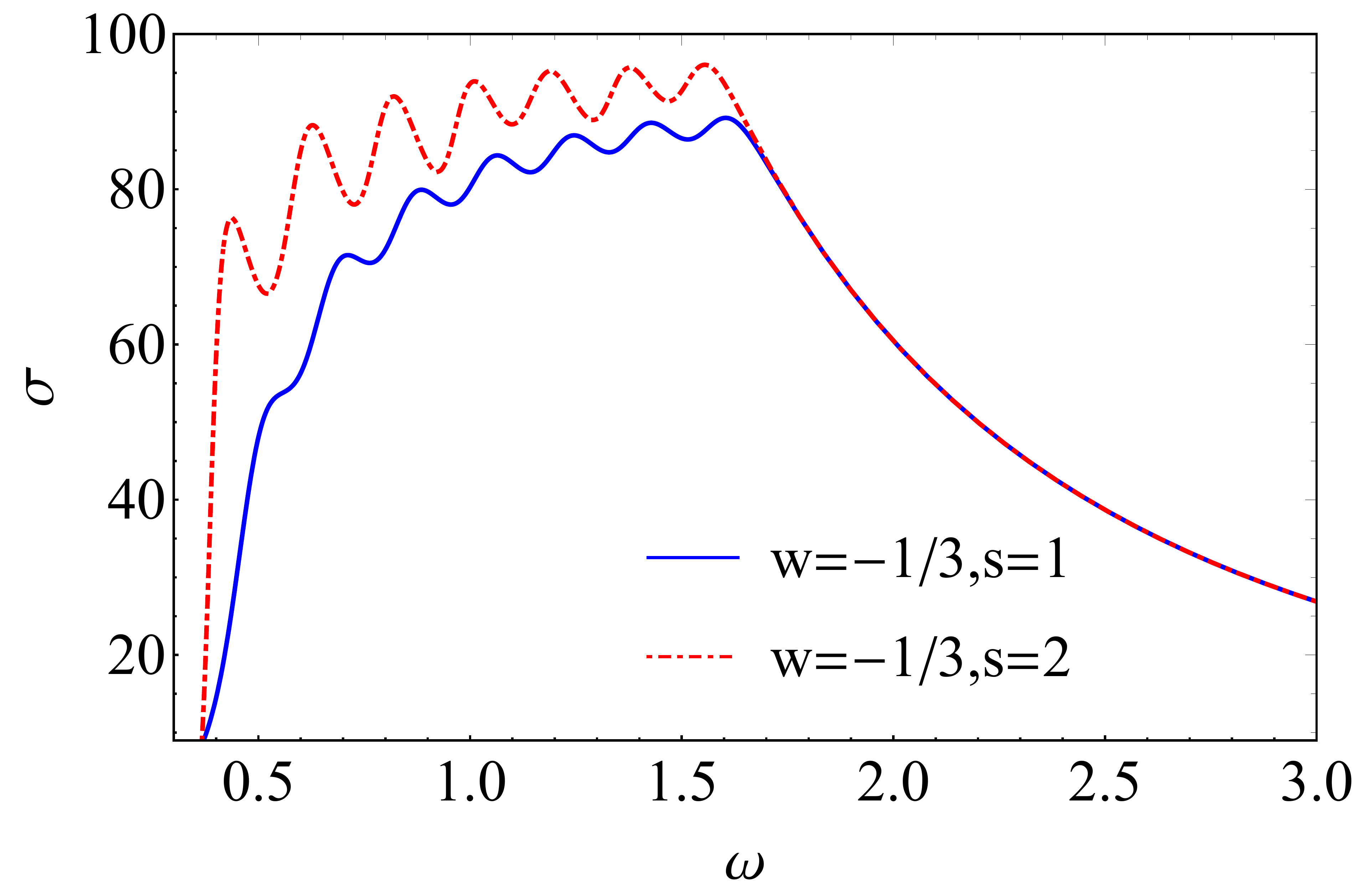}}
\centerline{(a)}
\end{minipage}
\hspace{7mm}
\begin{minipage}[t]{0.45\linewidth}
\centerline{\includegraphics[width=7.0cm]{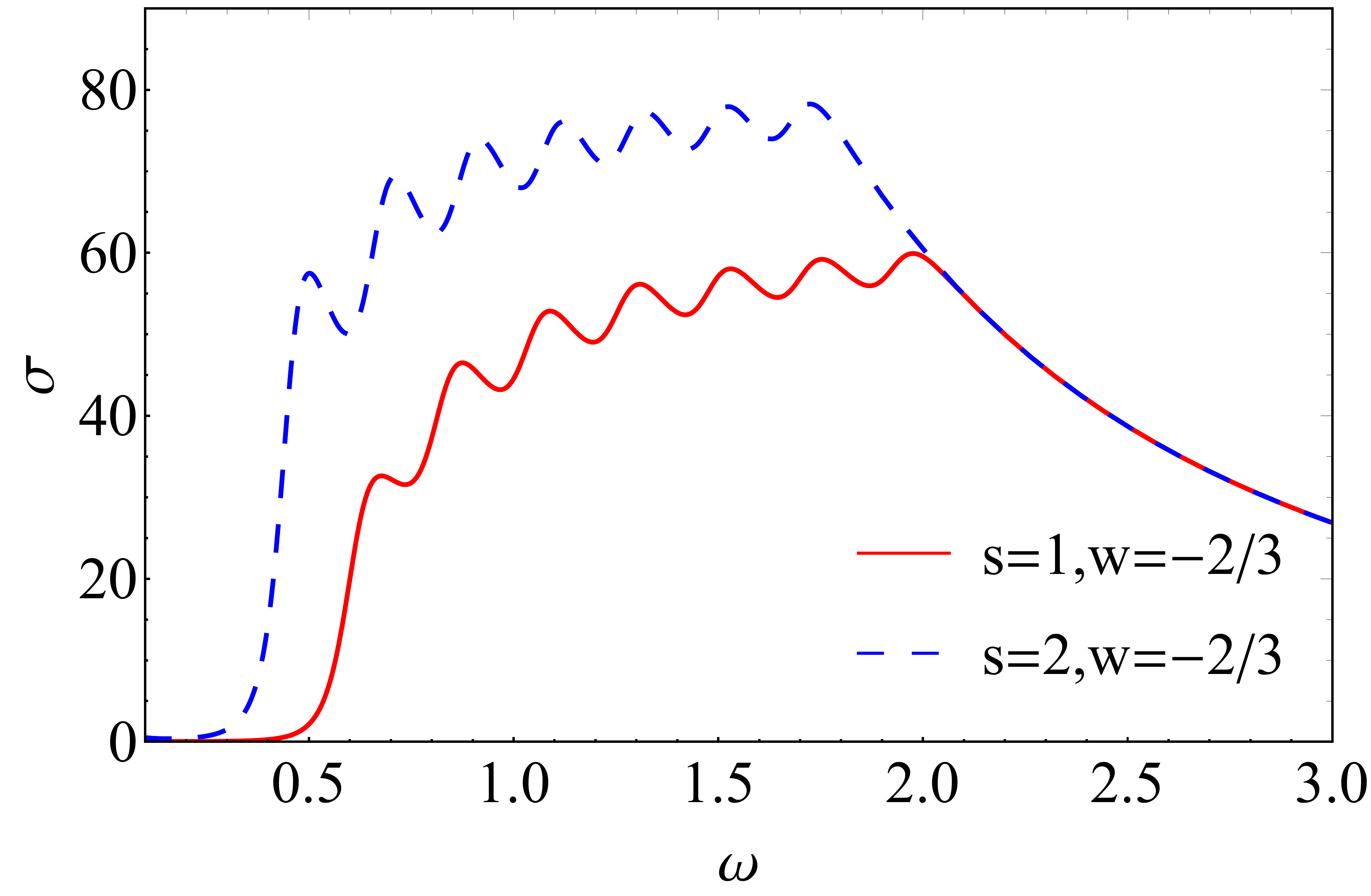}}
\centerline{(b)}
\end{minipage}
\end{tabular}
\caption{(a) Total absorption cross section($\sigma$) vs $\omega$ for $w=-1/3$, $M=1$, $q=0.25$,  $\lambda=0.0025$  and $c=0.02$; (b) Total absorption cross section($\sigma$) vs $\omega$ for $w=-2/3$, $q=0.5$, $M=1$, $c=0.001$ and $\lambda=0.0004$.}
\label{fig13}
\end{figure}

Fig.\ref{fig13} shows the total absorption cross section under electromagnetic ($s=1$) and gravitational ($s=2$) perturbations for Bardeen-Kiselev BH with cosmological constant, for convenience we have summed over $\ell=2$ to $\ell=8$ modes to determine $\sigma$. It shows that for a fixed frequency, absorption cross section is always larger for gravitational perturbations with respect to electromagnetic one and this result is consistent with the Ref.\cite{retSA}. Moreover, as the transmission coefficient approaches 1 at some critical value of $\omega$, whether it is regarding to the electromagnetic perturbations or gravitational perturbations and regardless of the BH parameters, the total absorption cross section falls off as $\frac{1}{\omega^{2}}$, therefore we could find the fall-off region in this figure.

{\centering  \section{conclusions} \label{secVI} }

In this work, we evoted to study the gravitational perturbations and electromagnetic perturbations for the Bardeen-Kiselev BH with cosmological constant in terms of quasinormal modes and compared it with RN-dSQ. We present the effective potentials under the two perturbations, it shows that both potentials are positive definite between the event and cosmological horizons, and they all have a single maxima, which indicates that for a fixed set of parameters, the potentials decrease with increasing parameters $\lvert w \lvert$ or $\lambda$, in other words, the smaller value of $\lvert w \lvert$ or $\lambda$ suppresses the emission modes for gravitational perturbation and electromagnetic perturbation. After that, by using the sixth order WKB method, we obtain the quasinormal frequencies of the Bardeen-Kiselev BH with cosmological constant and RN-dSQ under gravitational perturbations in Fig.\ref{fig2}, Fig.\ref{fig3} and Fig.\ref{fig4}, it shows that the response of Bardeen-Kiselev BH with cosmological constant and RN-dSQ in terms of the imaginary part of $\omega$ are different only when the charge parameter $q$ is varied. Besides, in Fig.\ref{fig3}, we can see that the absolute values of the imaginary parts as well as the real parts of the quasinormal frequencies of Bardeen-Kiselev BH with cosmological constant with quintessence $(c\neq0)$ are smaller compared to those without quintessence $(c=0)$, and as the parameter $c$ increases, they will decrease. Also from Fig.\ref{fig2} to Fig.\ref{fig4}, it shows that the oscillation frequency and the damping rate of Bardeen-Kiselev BH with cosmological constant increase with the increasing $w$, meanwhile, increasing $c$ or decreasing $w$ implies increasing the density of quintessence. Therefore, we can remark that due to the presence of quintessence, the gravitational perturbations of the Bardeen-Kiselev BH with cosmological constant damp more slowly and oscillate more slowly, and this behavior increases when increasing the density of quintessence, which implies that quintessence reduces the dissipative effect of the BH on its neighborhood.

From table \ref{tab1} to table \ref{tab3}, we calculated the quasinormal frequencies in electromagnetic perturbations of Bardeen-Kiselev BH with cosmological constant and compared frequencies with RN-dSQ, it shows that due to the presence of quintessence,  the electromagnetic perturbations around the Bardeen-Kiselev BH with cosmological constant damps faster and oscillates slowly, and this behavior is different with the gravitational perturbations. Therefore through these tabels, the response of Bardeen-Kiselev BH with cosmological constant and RN-dSQ with cosmological constant under electromagnetic perturbations are different when the charge parameter $q$, the state parameter $w$ and the normalization factor $c$ are varied.

Afterwards, regarding to the QNMs oscillation shape, we used the finite difference method to study the dynamical evolution of the nonlinear electrodynamics field perturbation and gravitational perturbation in the time domain.

Moreover, we studied the greybody factors of Bardeen-Kiselev BH with cosmological constant under gravitational perturbations and electromagnetic perturbations and observed that

1. The value of greybody bound is zero when the frequency is minimal, and the value of greybody bound turns out to be 1 when the frequency is large enough, which
implies that the wave is basically totally reflected when the frequency is small, while in view of the tunneling effect, part wave could pass through the potential barrier when the frequency increases, or the wave will not be reflected when the frequency reaches a certain level.

2. Under electromagnetic perturbations and gravitational perturbations, for a fixed frequency,  the response of greybody factor for different spacetime parameters are similar.

3. Due to the presence of quintessence, the transmission coefficient will increase.

Finally, we investigated the total absorption cross section under electromagnetic and gravitational perturbations for Bardeen-Kiselev BH with cosmological constant, it shows that for a fixed frequency, absorption cross section is always larger for gravitational perturbations with respect to electromagnetic one.

\begin{acknowledgments}
 This research was funded by the Guizhou Provincial Science and Technology Project(Guizhou Scientific Foundation-ZK[2022] General 491) and National Natural Science Foundation of China (Grant No.12265007).
\end{acknowledgments}

\end {document}